\documentclass[sigconf]{acmart}

\AtBeginDocument{%
  }

\setcopyright{acmlicensed}
\copyrightyear{2025}
\acmYear{2025}
\acmConference[IUI '25]{30th International Conference on Intelligent User Interfaces}{March 24--27, 2025}{Cagliari, Italy}
\acmBooktitle{30th International Conference on Intelligent User Interfaces (IUI '25), March 24--27, 2025, Cagliari, Italy}\acmDOI{10.1145/3708359.3712167}
\acmISBN{979-8-4007-1306-4/25/03}

\acmSubmissionID{6686}

\usepackage{diagbox}
\usepackage{multirow}
\usepackage{array}
\usepackage{listings}
\usepackage{xcolor}
\usepackage{transparent}
\usepackage{enumitem}
\usepackage[normalem]{ulem}

\definecolor{CharacterRed}{HTML}{F24822}
\definecolor{ItemGreen}{HTML}{14AE5C}
\definecolor{DialogueYellow}{HTML}{FFCD29}
\definecolor{ActionBlue}{HTML}{0D99FF}

\colorlet{characterColor}{CharacterRed!50}
\colorlet{itemColor}{ItemGreen!50}
\colorlet{dialogueColor}{DialogueYellow!50}
\colorlet{actionColor}{ActionBlue!50}

\begin{document}

\title{DancingBoard: Streamlining the Creation of Motion Comics to Enhance Narratives}


\author{Longfei Chen}
\authornote{Both authors contributed equally to this research.}
\orcid{0009-0002-4596-8093}
\affiliation{\institution{School of Information Science and Technology \\ ShanghaiTech University}
\city{Shanghai}
\country{China}}
\email{chenlf@shanghaitech.edu.cn}

\author{Shengxin Li}
\authornotemark[1]
\orcid{0009-0006-4117-3954}
\affiliation{\institution{School of Information Science and Technology \\ ShanghaiTech University}
\city{Shanghai}
\country{China}}
\email{lishx1@shanghaitech.edu.cn}

\author{Ziang Li}
\authornote{This work was done when Ziang Li was an undergraduate student at ShanghaiTech University.}
\orcid{0009-0005-5229-8325}
\affiliation{\institution{College of Design and Innovation \\ Tongji University}
\city{Shanghai}
\country{China}}
\email{ziangli@tongji.edu.cn}

\author{Quan Li}
\authornote{Corresponding Author.}
\orcid{0000-0003-2249-0728}
\affiliation{\institution{School of Information Science and Technology \\ ShanghaiTech University}
\city{Shanghai}
\state{Shanghai}
\country{China}}
\email{liquan@shanghaitech.edu.cn}

\renewcommand{\shortauthors}{Chen and Li et al.}
\newcommand{\clf}{\textcolor{black}}

\begin{abstract}
Motion comics, a digital animation format that enhances comic book narratives, have wide applications in storytelling, education, and advertising. However, their creation poses significant challenges for amateur creators, primarily due to the need for specialized skills and complex workflows. To address these issues, we conducted an exploratory survey (N=$58$) to understand the challenges associated with creating motion comics, and an expert interview (N=$4$) to identify a typical workflow for creation. We further analyzed $95$ online motion comics to gain insights into the design space of character and object actions. Based on our findings, we proposed \textit{DancingBoard}, an integrated authoring tool designed to simplify the creation process. This tool features a user-friendly interface and a guided workflow, providing comprehensive support throughout each step of the creation process. A user study involving $23$ creators showed that, compared to professional tools, \textit{DancingBoard} is easily comprehensible and provides improved guidance and support, requiring less effort from users. Additionally, a separate study with $18$ audience members confirmed the tool's effectiveness in conveying the story to its viewers.
\end{abstract}




\begin{teaserfigure}
    \includegraphics[width=\textwidth]{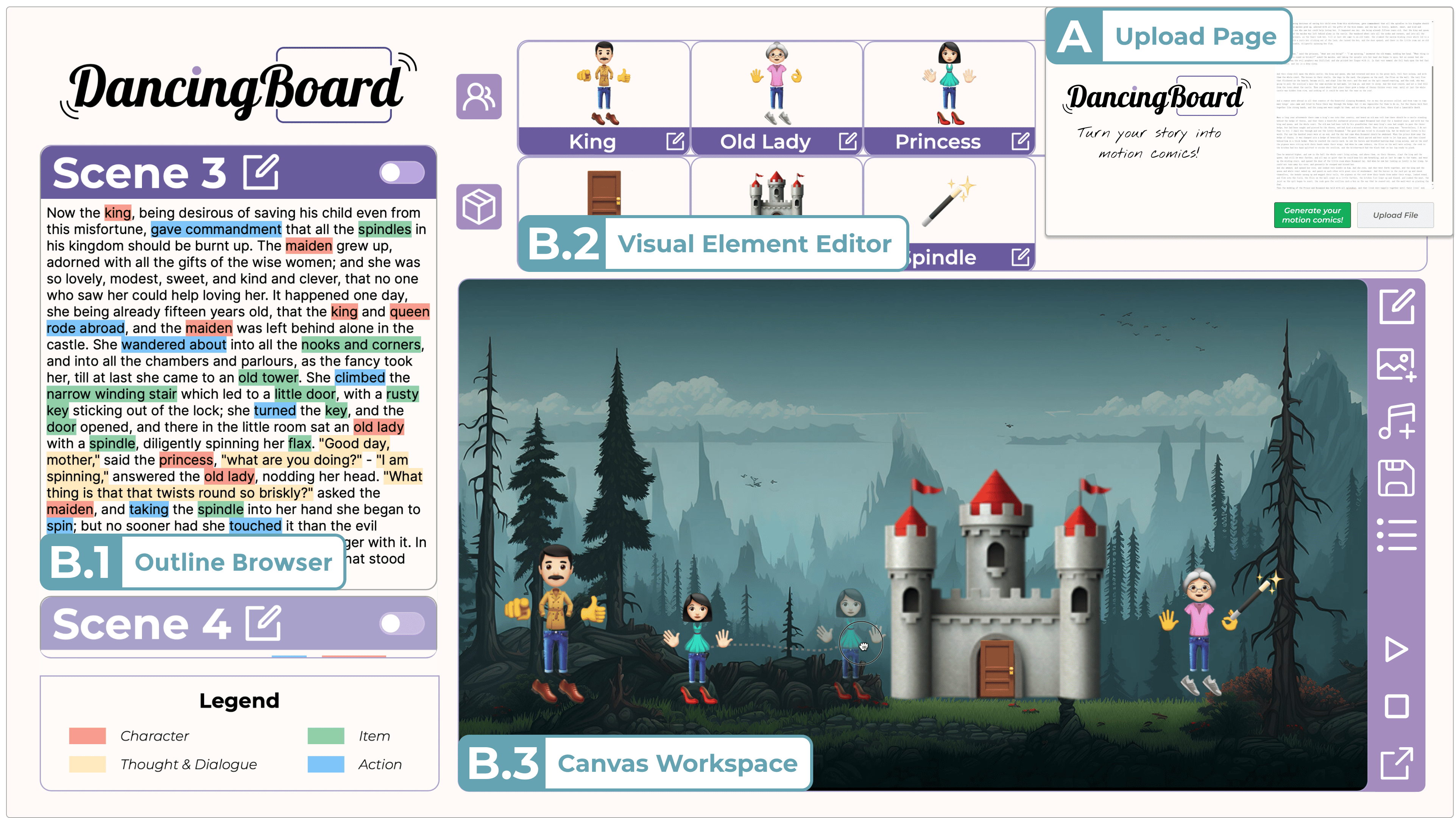}
    \caption[]{The \textit{DancingBoard} system comprises two primary pages: (A) the \textit{Upload Page}, enabling users to upload story text, and (B) the \textit{Creation Page}, offering core functionalities for creating motion comics. The \textit{Creation Page} encompasses three distinct views: (B.1) the Outline Browser, presenting the story content and scene outlines; (B.2) the Visual Element Editor, allowing previews and editing of visual elements like characters and objects; and (B.3) the Canvas Workspace, displaying the content and facilitating various editing operations.}
    \label{fig:overview}
\end{teaserfigure}

\maketitle

\section{Introduction}
\par
\begin{figure}[h]
    \centering
    \includegraphics[width=\linewidth]{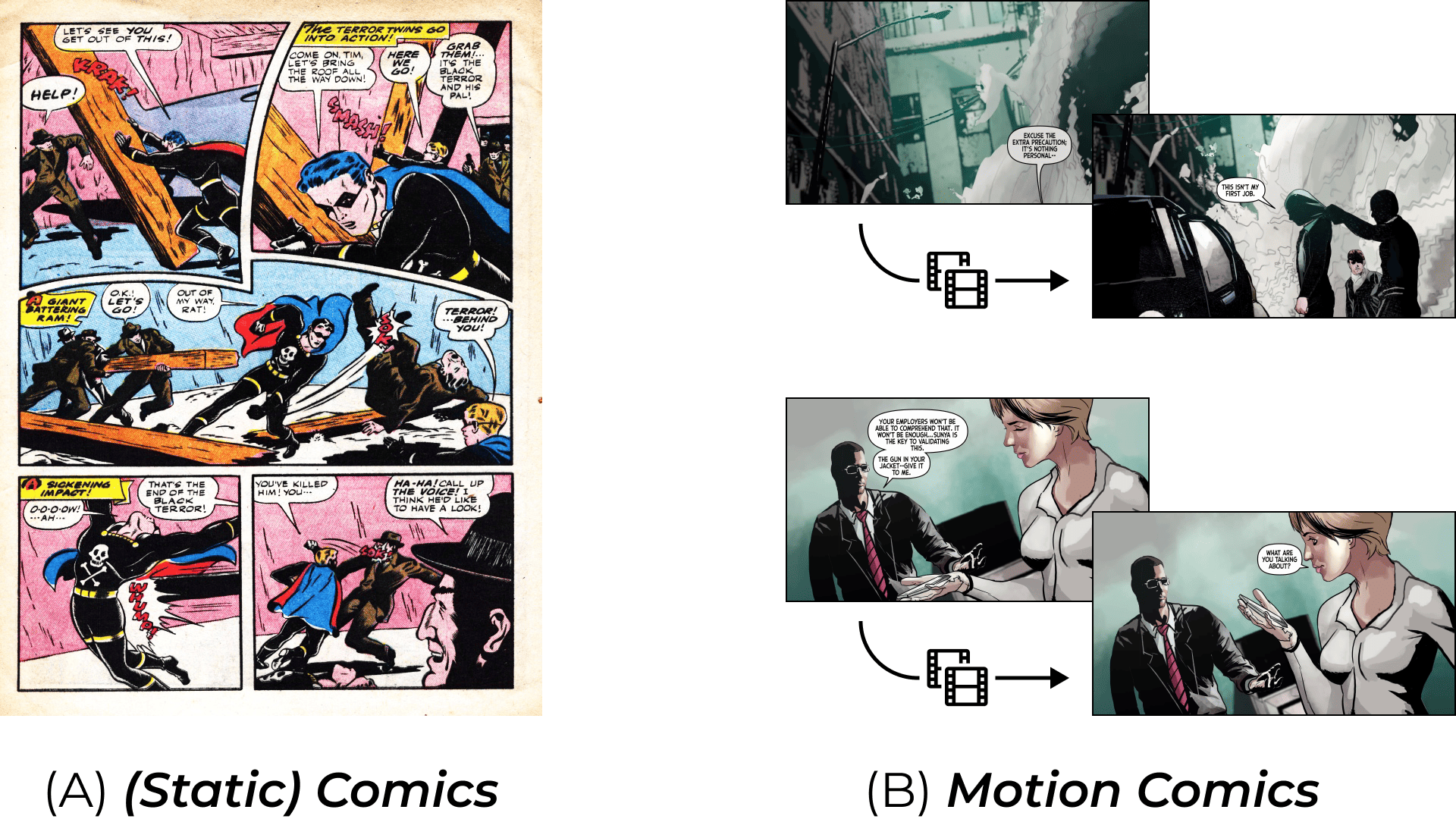} 
    \caption[]{A comparison between static comics ((A): \textit{The Black Terror \#11}~\cite{StaticMotionExample}) and motion comics ((B): \textit{The Accountant}~\cite{Accountant}) reveals a key difference in storytelling techniques. The adaptation of static comics to motion comics typically follows a ``Cinematic Approach''~\cite{smith2015motion}, where the presentation resembles animation or film. While static comics rely on panels and gutters (the spaces between panels) to convey the narrative, motion comics employ a full-screen canvas with animation effects applied to the visual elements, resulting in a more dynamic and immersive experience.}
    \label{fig:exampleMC}
\end{figure}

 Motion comics, a form of digital animation that ``\textit{appropriates and remediates an existing comic book narrative and artwork into a screen-based animated narrative}''~\cite{smith2012motion}, enhances the expressiveness of the original static comic by applying animation effects to the visual elements of the story~\cite{morton2015unfortunates}. Notable examples include \textit{Angry Birds: The Motion Comic}~\cite{AngryBirds}, \textit{A Night Out With SPIDER-MAN \& WOLVERINE}~\cite{ANightOutSpiderman} and \textit{The Accountant}~\cite{Accountant}. As shown in \autoref{fig:exampleMC}, unlike traditional static comics, motion comics can convey content within a single scene that would typically require multiple panels. Additionally, the transitions between panels are visually intuitive, significantly enhancing audience engagement~\cite{morton2015unfortunates}. 

\par With the development of online media, motion comics have been widely used in various fields such as storytelling ~\cite{morton2015unfortunates}, education~\cite{hashim2016theoretical, karlimah2021development} and advertising ~\cite{Onishi2020Method}. The popularity of short video platforms like \textit{TikTok} has made it easier for people to share their creations, further fueling the enthusiasm of amateur creators for producing motion comics~\cite{zeng2023exploring,wei2022historical,kapoor2018advances}. However, the process of creating motion comics typically requires specialized skills and tools. For instance, character design often relies on hand-painting or graphic design applications like \textit{Adobe Photoshop}~\cite{PS}, while animation effects in motion comics are achieved using professional tools such as \textit{Adobe After Effects}~\cite{AE}. \clf{Despite the rich functionality these tools offer, their complex interfaces and interactions place high demands on users' practice, requiring extensive effort to master. For amateurs, overcoming the steep learning curve can be particularly challenging. They often face numerous obstacles, such as figuring out how to get started in an unfamiliar field~\cite{chilana2010understanding,jahanlou2021challenges}, finding suitable learning resources~\cite{jahanlou2021challenges}, and effectively seeking help when problems arise~\cite{kiani2020would}.}

\par While research in media and Human-Computer Interaction (HCI) extensively covers creation methods~\cite{Onishi2020Method}, adaptation and media hybridization~\cite{smith2012motion,zeng2023exploring} and various applications of motion comics~\cite{hashim2016theoretical,karlimah2021development,Wulandari2022Feasibility}, there is still a lack of comprehensive understanding regarding the specific challenges encountered by amateurs during the creation process. To bridge this gap, we propose the first research question \textbf{(RQ1) What obstacles and challenges do amateur creators encounter when creating motion comics?} To investigate this question, we conducted an exploratory survey involving $58$ respondents, through which we identified five primary challenges encountered by amateur creators. First, professional tools have steep learning curves, requiring significant time and effort to master. Second, the lack of a typical workflow often leads to disorganized and inefficient creative processes. Third, collecting assets is time-consuming, as amateurs frequently search the internet for pre-existing materials. Though this requires fewer specialized skills, it remains a lengthy process. Fourth, amateurs struggle with portraying suitable animation effects due to limited experience. Lastly, editing and modifying visual elements or animations can be cumbersome and frustrating.

\par Existing tools and methods provide limited support in addressing the challenges outlined above. For example, some creation tools like \textit{Cartoon Animator}~\cite{CartoonAnimator} and  \textit{Synfig}~\cite{Synfig} simplify the creation process by introducing preset visual elements and binding them to controllable skeletal structures for easier animation effects. Although these tools technically lower the barriers to creating motion comics, they still do not provide users with adequate guidance. For users, these tools offer a range of controllable ``puppets'', but they may still encounter confusing questions, such as, ``\textit{What actions should I make these puppets perform to enhance their expressiveness?}'' Recent emerging text-to-video models such as \textit{Sora}~\cite{Sora}, \textit{Lumiere}~\cite{Lumiere} and \textit{Kling}~\cite{Kling} present alternative options. These tools can generate short video clips based on natural language descriptions, which may be further extended to motion comics. However, apart from a few exceptions, these tools are not yet widely publicly available. Furthermore, they have several limitations, particularly in terms of inconsistency in the generated content~\cite{zhang2023controlvideo, zhang2024towards}, difficulty in modifying the content~\cite{shin2024edit, zhang2024towards}, and lack of support for advanced animation techniques and personalized expressions~\cite{guo2023animatediff, zhao2025motiondirector}. Therefore, although the aforementioned methods offer some level of assistance, they still do not fully meet the needs of amateur creators. This further leads to our second research question \textbf{(RQ2) How to provide assistance for amateurs during their creation process?}

\par To address \textbf{RQ2}, we invited four domain experts for an online interview. The discussion focused on \textit{(1) the typical workflow for creating motion comics} and \textit{(2) potential support strategies for amateurs}. Based on the insights from the interviews, we identified four design goals and developed a motion comic creation tool named \textit{DancingBoard}. This tool features a concise interface design, follows a typical workflow, and offers users ample guidance and assistance at each step to minimize their workload and cognitive costs. Additionally, to alleviate the difficulties faced by amateurs in animation usage, we further explored \textbf{(RQ3) What is the design space of object actions in motion comics?} Specifically, we collected $95$ motion comic contents from online resources, extracted design patterns of character and object actions, and integrated these animation patterns as built-in suggestions into the \textit{DancingBoard} system. Subsequently, we evaluate the usability of the system through two research questions: \textbf{(RQ4) How effectively does \textit{DancingBoard} support the creation of motion comics?} and \textbf{(RQ5) Can the motion comics created using \textit{DancingBoard} accurately convey stories to the audience?} To investigate \textbf{RQ4}, we conducted a user study with $23$ amateur creators, comparing our system with the other existing tool. The results indicated that \textit{DancingBoard} offers an intuitive and easy-to-learn workflow, demonstrating advantages in improving creative efficiency and reducing cognitive workload. To explore \textbf{RQ5}, we conducted another user study involving $18$ participants. In this experiment, we gathered feedback from participants after viewing both the motion comics created with \textit{DancingBoard} and the ones created with professional tools. The results showed that the motion comics produced with \textit{DancingBoard} effectively conveyed the story content to the audience in a simple yet accurate visual style. In summary, the contributions of this study are as follows:
\begin{itemize}
    \item We conduct a formative study to identify the challenges faced by amateur creators and outline a typical workflow for creating motion comics.
    \item We summarize the design space for character and object actions in motion comics, offering a structured approach to authoring.
    \item We develop \textit{DancingBoard}, an integrated authoring tool designed to streamline the motion comic creation process for amateur creators.
    \item We validate the usability of our approach through two user studies, demonstrating the effectiveness of the proposed tool.
\end{itemize}

\section{Related Work}
\par The process of transforming textual stories into motion comics can be divided into two key phases. The first phase involves comprehending the story content, segmenting the narrative into individual frames, and generating visual illustrations for each frame to create static comics. The second phase focuses on animating these frames and implementing transitions to convert the static comics into motion comics. In this section, we will first provide an in-depth overview of the existing tools and workflows used for motion comic creation. Following that, we will discuss related works pertaining to each of the two phases in detail.

\subsection{Motion Comic Authoring Tools}
\label{sec:2.1}
\par Traditionally, the creation of professional motion comics begins with a blank canvas. Authors first design static visual elements, typically by manually drawing them using software platforms such as \textit{Adobe Illustrator}~\cite{AI}, \textit{Adobe Photoshop (PS)}~\cite{PS}, or \textit{Sketch}~\cite{Sketch}. Next, they use tools like \textit{Adobe After Effects (AE)}~\cite{AE} or \textit{Maxon Cinema4D}~\cite{C4D} to add animation effects to these static elements, bringing them to life. While these professional tools offer extensive functionality, they also come with a steep learning curve and high technical barriers, requiring significant time and practice to master. Therefore, some authoring tools, such as \textit{Cartoon Animator}~\cite{CartoonAnimator} and \textit{Synfig}~\cite{Synfig}, have introduced features like skeleton systems and camera controls to simplify the creation of motion comics. \clf{However, these approaches still pose cognitive challenges for users, as they need to grasp unfamiliar domain concepts such as layer management and animation curves~\cite{chilana2010understanding}. Moreover, they must go through the time-consuming and often non-immediately rewarding ``pre-production phase'', which involves tasks like transforming project briefs into scripts, creating storyboards, and editing prototypes~\cite{jahanlou2021challenges}.}

\par To lower the barrier to creation, researchers in the HCI field have developed simplified tools to help users more easily add animations to visual elements. For instance, \textit{Draco}~\cite{Draco} introduced a sketch-based interface that allows users to create rich animations for their drawings. \textit{Kitty}~\cite{Kitty} further enables users to add animations and define chained causal effects between visual subjects. Willett et al.~\cite{nora2018interativetool} presented an interactive tool that animates visual elements of a static picture using simple sketch-based markup. \clf{In addition, general-purpose visual storytelling animation tools, such as slide-creation tools (e.g., \textit{PowerPoint}~\cite{PPT}) and web-based applications (e.g., \textit{Rive}~\cite{Rive}, \textit{Lumen5}~\cite{Lumen5}, \textit{Animaker}~\cite{Animaker}), can also support the creation of simple animations. The related studies have been comprehensively reviewed and summarized in \cite{jahanlou2023end}.}

\par It is evident that current motion comic authoring tools represent two extremes. On one end are professional-grade software solutions, which offer a wide range of features and high levels of customization, but have very steep learning curves that require users to master specific technical knowledge or skills. On the other end are tools that simplify the creation process, but their ability to support creative work remains limited. It is worth mentioning that the rapid development of Artificial Intelligence Generated Content (AIGC) technology provides potential opportunities for creating motion comics and bridging the gap between these two extremes. \clf{On one hand, text-to-image tools (e.g., \textit{Krock}~\cite{Krock}, \textit{Boords}~\cite{Boords}, and \textit{Storyboarder}~\cite{StoryboarderAI}) simplify the process of creating visual assets. On the other hand, text-to-video tools (e.g., \textit{Sora}~\cite{Sora}, \textit{Lumiere}~\cite{Lumiere}, \textit{Kling}~\cite{Kling}, and \textit{Anim-Director}~\cite{Anim-Director}) provide a more streamlined solution, allowing users to generate dynamic video content with just text input. While these LLM-based tools significantly facilitate the creation of motion comics, their limitations cannot be overlooked. First, the output generated by LLMs is not always consistent for identical textual descriptions. This inconsistency is particularly evident when dealing with ambiguous or multi-interpretational text, which can lead to results that deviate from the user's intents~\cite{maddigan2023chat2vis, zhang2023controlvideo, zhang2024towards}. Second, the content generated by these tools often lacks flexibility in terms of editing capabilities, making it challenging for users to refine and optimize the results~\cite{shin2024edit, zhang2024towards}. This limitation is especially critical in the context of motion comic creation, where users often need to fine-tune visual layouts, animation pacing, and narrative structures to ensure both coherence in storytelling and visual appeal.}

\clf{In this study, we conducted an in-depth investigation into the challenges faced by amateur creators in producing motion comics and developed a tool called \textit{DancingBoard} to address these issues. \textit{DancingBoard} takes text as the primary input and guides users step-by-step through each stage of motion comic creation, following the general paradigm of text-driven animation creation~\cite{ASAS,sumi2006animated,DancingWords,DataParticles,DrawTalking,DataPlaywright}. Compared to traditional methods, our system lowers the technical barriers to entry and enhances the creative experience for amateur users through a modular design approach. Specifically, the system features an intuitive interface that allows users to flexibly adjust narrative scripts, visual assets, and scene designs. Additionally, \textit{DancingBoard} integrates predefined animation effects, enabling even users with no prior animation experience to quickly experiment with different creative combinations, thus reducing workload and cognitive burden.}

\subsection{Framing the Narrative Text}

\par Comics and motion comics serve as expanded visual interpretations of narrative texts, enriching their expressive qualities. Yet, as observed by Evangelia et al.~\cite{Evangelia2018Adapting}, comics created by different individuals can differ significantly even based on the same story. This could pose similar challenges when creating motion comics: while each creator may infuse their own perspectives and aesthetic choices into their work, straying too far from the essence of the original narrative text can potentially result in misunderstanding or confusion among readers.

\par To address this issue, we draw inspiration from previous research. For example, Ueno Miki conducted structural analysis on four-scene comic stories, where the story was analyzed in terms of settings, characters, plots, actions and dialogue ~\cite{ueno2019structure}. Francis Glebas ~\cite{glebas2012directing} transformed original story text into structured scripts, a specific form of film production including dialogue, actions, time, and scene descriptions. Similarly, approaches such as ~\cite{das2023storytube, zhang2021write, zhang2019generating} have utilized narrative texts to guide animation creation. Building upon these approaches, we employed automated methods to frame the narrative text. Specifically, the framing process consists of two consecutive steps: (1) \textit{segmenting the story text into different scenes based on environmental changes}, and (2) \textit{extracting key information from these segmented scenes}, including characters, objects, actions, dialogue, and thoughts. For tasks involving natural language understanding, we leveraged the state-of-the-art LLM, GPT-4o, and meticulously designed prompts. Compared to rule-based methods ~\cite{manning1999foundations} and statistical learning methods ~\cite{mccallum1998comparison,anandika2021review}, LLMs demonstrate superior comprehension of overall semantics and contextual information ~\cite{moghaddam2023boostingtheoryofmindperformancelarge}, thereby offering more precise and adaptable narrative text parsing capabilities.

\subsection{Animation in Visual Storytelling}

\par Animation plays an increasing important role in storytelling. In contrast to static text and images, animated content excels in stimulating visual senses and captivating audience attention ~\cite{kosara2013storytelling,bradley1992remembering}. Chevalier et al. ~\cite{chevalier2016animations} delved into $23$ distinct roles that animation fulfills in user interfaces, organized into five categories where one of them is \textit{``storytelling and visual discourse''}.

\par Previous research on animation in visual storytelling within the HCI community has predomicantly focused on the domain of data storytelling. For example, Thompson et al. ~\cite{thompson2020understanding} analyzed animation transitions across dimensions such as objects, graphics, data, and timing, identifying ten transition types. Shi et al. ~\cite{shi2021communicating} explored the interplay between animation and visual narration in data videos, synthesizing four animation techniques and eight visual narrative strategies from $82$ examples. Ouyang et al.~\cite{ouyang2024noteplayer} analyzed $38$ tutorial videos and identified eight categories of visual narratives and six types of creator behaviors, exploring the relationships between the two. Amini et al. ~\cite{amini2018hooked} discovered that animation and pictograms can heighten viewer engagement when watching data videos.

\par \clf{While the aforementioned studies primarily explore the use of animation to express and communicate data, the creation of motion comics presents unique challenges, as creators often focus on designing animations to depict entities and their actions. For example, creators may face questions like ``How do I animate the action `\textit{Mom passes the kitchen}' or \textit{Mom gives Mary an apple}'?''—tasks that can feel particularly overwhelming for inexperienced amateurs. Therefore, we collected and analyzed $95$ motion comics, identifying $8$ common animation patterns in motion comics. These patterns were subsequently integrated into \textit{DancingBoard} as predefined animation effects, offering targeted support for creators during the creation process.}

\section{Formative Study}
\label{sec:formative_study}

\par To explore \textbf{(RQ1) What obstacles and challenges do amateur creators encounter when creating motion comics} and \textbf{(RQ2) How to provide assistance for amateurs during their creation process}, we initiated a formative study. We first conducted an exploratory survey (\autoref{sec:Exploratory}) to understand the skill levels of amateurs, the tools they use to create motion comics, and the potential challenges they might encounter. Furthermore, to establish a typical workflow for creating motion comics and explore entry points for assisting amateur creators, we organized an online interview (\autoref{sec:3.2}) with four domain experts. Finally, based on the findings from the exploratory study and feedback from the interview, we summarized a set of design goals (\autoref{sec:3.3}).

\subsection{Exploratory Survey}
\label{sec:Exploratory}

\subsubsection{Survey Protocol} We designed a survey questionnaire using \textit{Microsoft Forms} and distributed it through social networks to invite respondents to participate. In addition to collecting demographic information, the questionnaire also explored respondents' experiences and skills in creating motion comics. Subsequently, it further inquired about their workflows and the tools they used in actual creation. \clf{To better understand the challenges they faced, three authors brainstormed potential pain points based on their own experiences and existing literature~\cite{jahanlou2021challenges,smith2015motion,DataParticles,ouyang2024noteplayer}. These pain points were included in the questionnaire, such as ``\textit{the need to master complex professional tools and skills}'' and ``\textit{difficulty in following a clear workflow during creation}''.} Respondents were asked to rate these pain points using a standard 5-point Likert scale ($1$ -- strongly disagree, $5$ -- strongly agree). Additionally, the questionnaire included open-ended questions to encourage respondents to share any difficulties they had encountered.

\subsubsection{Respondents} We received a total of $58$ valid responses (referred as \textbf{P1--58}, comprising $13$ females, $39$ males and $6$ individuals whose gender was unspecified). Of these, $30$ were undergraduates, $26$ were pursuing master's degrees, and $2$ were pursuing Ph.D. degrees. Their age spanned from $17$ to $28$ years, with an average of $22.4$ ($SD = 3.37$). \clf{Notably, all respondents had experience in creating motion comics and were unaware of the authors’ identities while completing the questionnaire, as it was distributed anonymously.}

\begin{figure*}[h]
    \includegraphics[width=0.9\linewidth]{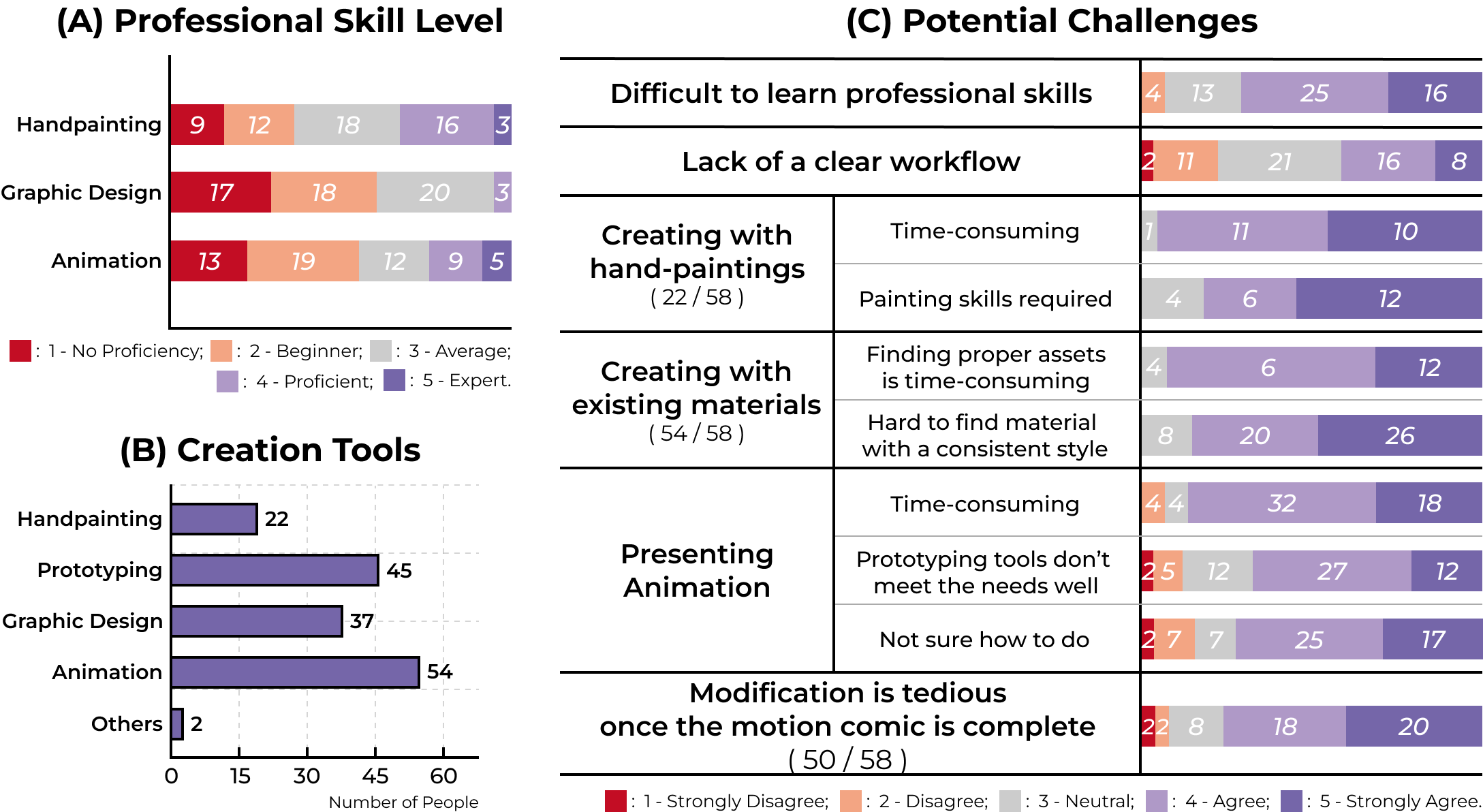}
    \caption[]{The findings from the exploratory survey touch upon three main aspects: (A) proficiency levels in acquiring professional skills, (B) frequently used tools for motion comics creation, and (C) potential challenges encountered during the creation process.}
  \label{fig:survey_result}
\end{figure*}

\subsubsection{Results and Findings} Based on results as shown in \autoref{fig:survey_result}, we identified the following key findings (\textbf{F.1}--\textbf{F.5}).

\par \textbf{F.1. The learning curve for professional tools is steep, with interface prototyping tools being favored for drafting.} The majority of respondents exhibit some deficiency in professional skills (\autoref{fig:survey_result} (A)). When queried about the tools employed for creating motion comics, $93.1\%$ ($54/58$) mentioned professional tools such as \textit{Adobe After Effects} and \textit{Cartoon Animator}. Responses show that mastering these tools pose significant challenge for novices. \textbf{P8} noted, ``\textit{Basically these (professional tools) are the only choices for creating decent motion comics ... Mastering them is daunting. I have to learn a series of complicated concepts.}'' Notably, $77.6\%$ ($45/58$) mentioned the need of creating drafts for exchanging ideas and evaluating designs, with interface prototyping tools such as \textit{Figma}~\cite{Figma} being widely used for their built-in easy-to-use animation effects. \textbf{P20} noted, ``\textit{It's too costly to remake the creation, so we make simplified prototypes using tools such as \textit{Figma} and show them to others to receive responses. It's much easier for me to navigate these transition effects, since I've used it a lot during my previous studies.}'' 

\par \textbf{F.2. Lack of a clear workflow.} Upon requesting each participant to outline their creation process briefly, it became evident that the majority lacked a structured approach. Many described their approach as ``\textit{improvised}''. As shown in \autoref{fig:survey_result} (C), when queried about the challenges related to ``\textit{not being able to clearly grasp the workflow}'', $77.6\%$ ($45/58$) of respondents rated this agreement level at $3$ (Neutral) or higher (Agree or Strongly Agree).

\par \textbf{F.3. Substantial time is spent on sourcing assets.} A notable $93.1\%$ ($54/58$) of respondents acknowledged utilizing pre-existing assets in their motion comics creation process, for example, visual props and background music (BGM). These assets are typically sourced from various search engines and websites. For instance, \textbf{P9}, \textbf{P30} and \textbf{P47} favor \textit{Pinterest}~\cite{Pinterest}, while \textbf{P7} and \textbf{P31} mention \textit{Digital Comic Museum}~\cite{DCM}. \textbf{P30} elaborated, ``\textit{I'm really not good at drawing, so using pre-made visual elements is my go-to alternative.}'' \textbf{P15} commented, ``\textit{It's impossible for us designers to create our own BGM.}'' The issue of time consumption also persists (\autoref{fig:survey_result} (C)). Consensus among respondents is that finding suitable assets is challenging, demanding a significant investment of time and effort.

\par \textbf{F.4. Feel uneasy to present the animation.} More than $84.5\%$ ($49/58$) respondents admitted to feeling unsure about how to present the animation effects in motion comics. \textbf{P23} elaborated, ``\textit{Character actions, background transitions ... I might have some vague ideas, but I just don't know how to make them happens.}'' As previously mentioned, interface prototyping tools are the preferred option (\autoref{fig:survey_result} (B)). However, when their built-in animation features fail to meet users' needs, they find themselves at a loss.

\par \textbf{F.5. Modification and revision is a tedious task.} A substantial $86.2\%$ ($50/58$) of respondents highlighted the necessity for frequent adjustments to visual elements or interactions during the creation process. Many ($46/50$) among them acknowledged that these modifications can be quite time-consuming. \textbf{P32} commented, ``\textit{Small tweaks can mess with other parts (of the motion comic), and then you end up having to make even bigger changes.}''

\subsection{Expert Interview}
\label{sec:3.2}
\par During our exploration survey, we found that amateurs lack a clear and typical workflow to guide them and encounter various difficulties during their practice. Thus, to achieve two objectives: \textbf{(1) identifying a typical workflow for creating motion comics} and \textbf{(2) devising better support strategies for amateurs}, we engaged four domain experts (\textbf{E1--4}) in an online interview.

\subsubsection{Expert Team} 
\par The expert team (\textbf{E1--4}) comprises individuals from our collaborating universities and institutions. All members of the team hold formal degrees in design, art, or digital media art, and possess over five years of experience in producing visual content. Additional detailed demographic information is available in \autoref{tab:experts}.

\begin{table}[h]
  \centering
  \caption{Demographic information about experts, including gender identity, range of age, occupation, and experience of motion comics authoring in years (denoted as Exp.).}
  \label{tab:experts}
  \begin{tabular}{clclc}
    \toprule
    \textbf{ID} & \textbf{Gender} & \textbf{Age} & \textbf{Occupation} & \textbf{Exp.} \\
    \midrule
    \textbf{E1} & Female & $35$ -- $39$ & Storytelling Director      & $12$ \\
    \textbf{E2} & Female & $25$ -- $29$ & Digital Media Art Instructor & $9$  \\
    \textbf{E3} & Male   & $25$ -- $29$ & Motion Comics Animator & $6$  \\
    \textbf{E4} & Female & $20$ -- $24$ & Motion Comics Animator   & $5$  \\
    \bottomrule
  \end{tabular}
\end{table}

\subsubsection{Interview Procedure} 
\par During the interview, experts were initially queried about their customary workflow of creating motion comics. Subsequently, they were encouraged to share any collaboration or mentoring experiences they had encountered with amateur creators. Following this, we presented the results and insights gleaned from our exploratory study. The subsequent discussion revolved around the challenges amateur creators encounter, their expressed needs, and strategies for offering relevant support. The interview took place via the online platform \textit{Zoom Meeting}, spanning approximately two hours in duration. With participants' consent, both the audio recording and screen sharing were captured and archived.

\subsubsection{Typical Workflow of Motion Comic Creation}
\label{sec:workflow}
\par The experts disclosed their professional authoring workflow, which we subsequently formalized into a six-step pipeline as follows:

\par \textbf{Step 1:} Conduct a comprehensive analysis of the story content, extracting logical elements and segmenting them into scenes. Experts typically employ a two-column script during this phase, wherein the first column dissects the textual narrative into multiple components, and the second column showcases the corresponding visual effects to be crafted.

\par \textbf{Step 2:} Design the visuals for pivotal scenes, characters, props, and other key elements within the story. Depending on specific requirements, the design can range from simplified sketches (created free-hand) to high-fidelity prototypes generated using professional tools such as \textit{Photoshop}, \textit{Illustrator} and others. Notably, some of the visual elements are created in multiple versions, allowing for applying animation effects in the following steps.

\par \textbf{Step 3:} Illustrate static frames based on the visual design prototypes. It involves managing the layout of each visual elements such as characters and speech bubbles on the canvas. This process ensures that every frame accurately portrays the narrative progression, serving as a guide for subsequent production stages.

\par \textbf{Step 4:} Apply animation effects to each static frame. Each visual element is treated as an independent layer and is arranged on the animation timeline. Manipulating the layers on the timeline by turning them on and off, transforming them, and adding ``tweening'' to create smooth transitions between frames forms the basic movements of static elements. For more complex character animations, \textit{Adobe Character Animator} is used to rig facial features and lip-sync, with added physics for natural motion. Finally all elements are brought together in \textit{After Effects}, enhanced with visual effects, layered movements, and spatial depth.

\par \textbf{Step 5:} Add audio elements. Instead of using speech balloons for presenting character dialogues or other sounds, some creators prefer using voice-overs and soundtrack scores. It's also possible for combining speech balloons and soundtracks. Therefore, audio elements are treated as independent layers and arranged on the animation timeline. 

\par \textbf{Step 6:} Refine and iterate as necessary. Creators are required to refine the content of the outcome based on the feedback from peers and clients. \textbf{E3} remarked, ``\textit{It can be pretty tough, especially when you are dealing with all sorts of feedback from tricky clients. Rework is too common in the industry.}''

\subsection{Design Goals}
\label{sec:3.3}

\par After gathering insights and suggestions from experts regarding the challenges and needs articulated by amateur creators in the exploratory survey, we organized the information and derived the following five design goals.

\par \textbf{[DG1] Reduce complexity and learning curve.} All four experts emphasized that professional authoring demands a high level of proficiency in drawing or design skills, often necessitating the use of advanced tools. However, for amateurs, mastering these professional skills and tools poses a significant challenge (\textbf{F.1}). \textbf{E3} noted, ``\textit{We can't really expect someone who's not a pro to nail these tools like a pro would. I mean, when I was a novice creator, I went through the same struggles. Clearly it would be harder for amateurs.}'' Consequently, he recommended that our approach should prioritize reducing the reliance on professional skills while still providing control over the steps of authoring motion comics.

\par \textbf{[DG2] Provide guidance following typical workflow.} \textbf{E1} highlighted the advantages of adhering to a typical workflow, noting that it aids amateurs in clarifying their thoughts and conducting their work more effectively. She remarked, ``\textit{A lot of novice creators tend to get stuck on too detailed aspects when planning, which would waste too much time. Overwhelmed by that much details, they are unable to take the first step. This reflects their unfamiliarity with the workflow. Therefore, why not provide them some good guidance? This could help them manage their thoughts and kick-start their work, and finally boosting their quality and efficiency.}''

\par \textbf{[DG3] Facilitate access to visual elements.} Visual elements play a crucial role in motion comics. However, as addressed in the findings, sourcing visual assets takes up too much of amateur creators' time (\textbf{F.3}) while crafting visual elements with professional tools proposes challenge for them (\textbf{F.1}). As a workaround, many resort to utilizing prefabricated content from online asset libraries, but sourcing and modifying such content can be challenging. Experts proposed integrating existing asset libraries into the system and implementing categorization methods to facilitate easier access to visual content for users.

\par \textbf{[DG4] Lower barriers to presenting actions through pre-defined animations.} Amateurs often encounter significant challenges when attempting to present actions in motion comics (\textbf{F.4}). As \textbf{E4} pointed out, ``\textit{Even experienced creators need a ton of practice to really good at it, and it would only be harder for beginners.}'' While interface prototyping tools such as \textit{Figma} provides features such as \textit{Smart Animate}, their animation effects remain stiff and are mainly designed for showcasing UX, which are not entirely suitable for motion comics. Hence, we aim to condense the realm of actions and translate it into a set of predefined animations designed for motion comics, thereby simplifying the process and lowering barriers to incorporating interactive elements into the story.

\section{Design Space for Motion Comics}
\label{sec:designspace}

\par \clf{Based on previous research~\cite{kuhlman2020design,groensteen2007system,ComicsResearchInComputerScience,laubrock2020computational,smith2015motion}, we identified the following key elements for motion comics: object (mainly characters and items), backgrounds, text, and audio. Among these, backgrounds, text, and audio exhibit a relatively limited design space. Backgrounds are typically presented as images or utilize blurring effects to create a sense of spatial depth~\cite{smith2015motion}. Text serves as an important medium for conveying dialogue, thoughts, or other narrative content, usually accompanied by speech bubbles~\cite{smith2012motion, smith2015motion}. Audio can be roughly categorized into two types: BGM and character voice-overs. BGM is commonly used to enhance the atmosphere, while character voice-overs express the character's lines, contributing to a stronger sense of immersion~\cite{smith2015motion}.}

\par \clf{In contrast, the design space for object actions is much broader, as action requires various forms of animated expression based on specific behaviors. This presents a challenge for amateur creators when depicting object actions, as we observed in \autoref{sec:formative_study}. Therefore, we posed \textbf{(RQ3) What is the design space of object actions in motion comics?} to explore how to address this challenge effectively. To tackle this issue, we collected $95$ motion comic examples, aiming to identify common design patterns within them. Our approach includes several steps: first, sourcing high-quality motion comic materials from online platforms; second, analyzing the material library from two dimensions; and finally, collaborating with two animation experts to co-construct the design space.}

\subsection{Data Collection}
\par To ensure the quality and representativeness of the collected data, we established three inclusion criteria: (1) adherence to storytelling structure; (2) portrayal of at least one object (character or item); and (3) dynamic content utilizing animation techniques. Adhering to these criteria, we curated a dataset of $95$ motion comics from various online platforms such as \textit{YouTube} and \textit{TikTok}. Keywords examples are ``Motion Comics'' and ``Animated Comic Book''. The total duration of the collected videos amounted to $550$ minutes, ranging from as short as $57$ seconds to as long as $16$ minutes and $53$ seconds, and a mean duration of $5.79$ minutes. All content underwent thorough review to ensure compliance with the established inclusion criteria.

\subsection{Data Analysis}

\par Our analysis of the material library focused on two main dimensions. First, we examined the atomic animation operations utilized in the motion comics, and second, we scrutinized the classification of object actions depicted within them.

\subsubsection{Dimension I: Atomic Animation Operation} In motion comics, diverse animations are frequently utilized to amplify their expressive power. Yet, comprehending and proficiently executing complex animation techniques can prove challenging for novice amateurs. Hence, we aim to dissect animation techniques found in motion comics through the lens of atomic operations. Breaking down animations into their smallest manageable units (i.e. \textit{atomic operations}) enables us to precisely characterize their structure and implementation methods, thereby enhancing comprehension, especially for amateurs.

\par To establish the appropriate categorizations of atomic animation operations, we conducted a survey of literature on animation design~\cite{parent2012computer,johnston1981illusion,williams2012animator} and examined built-in animation effects in existing tools~\cite{Figma,AE,CartoonAnimator}. Based on our findings, we propose the following six fundamental atomic operations.

\par \raisebox{-1.3ex}{\includegraphics[height=4ex]{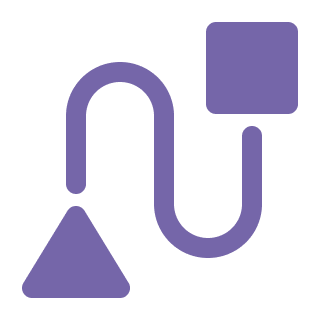}} \textbf{Path Movement}: Path movement refers to the action of guiding an object along a designated trajectory. This trajectory may take the form of a straight line, a curve, or a customized path. Utilizing path movement can enhance the audience's experience by creating a sense of natural fluidity, particularly when illustrating the movement of an object.

\par \raisebox{-1.3ex}{\includegraphics[height=4ex]{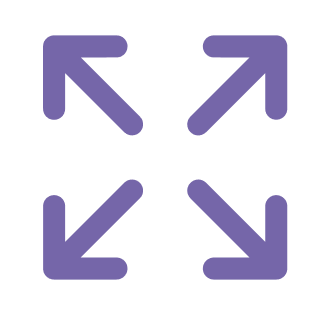}} \textbf{Scale}: Scaling refers to adjusting the size ratio of an object. This adjustment can involve enlarging or reducing the object, thereby altering its size within the frame. Scaling is a versatile technique that can be employed to emphasize the importance of an object, indicate its proximity or distance, or achieve a gradual fading effect.

\par \raisebox{-1.3ex}{\includegraphics[height=4ex]{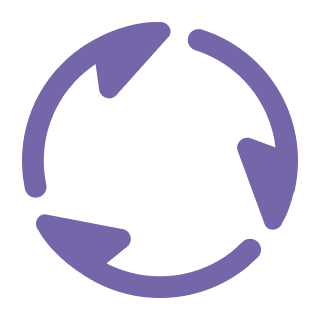}} \textbf{Rotation}: Rotation involves turning an object around a designated center point. This movement can occur either clockwise or counterclockwise, and it offers the flexibility to adjust the speed and angle of rotation. Rotation is commonly used to infuse objects with dynamism, capture attention, or produce distinctive visual effects.

\par \raisebox{-1.3ex}{\includegraphics[height=4ex]{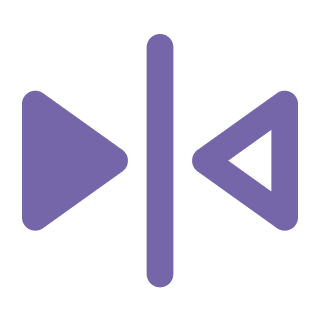}} \textbf{Flip}: Flipping involves creating a mirrored image of an object either horizontally or vertically. Horizontal flipping reverses the object from left to right, while vertical flipping flips it from top to bottom. Flipping is commonly used to change the orientation of an object or achieve a symmetrical effect.

\par \raisebox{-1.3ex}{\includegraphics[height=4ex]{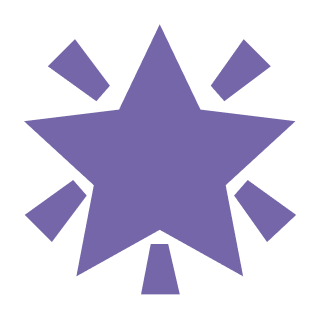}} \textbf{Appearance}: Appearance refers to the gradual emergence of an object on the screen from an initially invisible state. Objects can transition from transparent to full visibility or progressively enlarge in size. Appear animations are frequently employed to introduce new objects or smoothly transition between scenes.

\par \raisebox{-1.3ex}{\includegraphics[height=4ex]{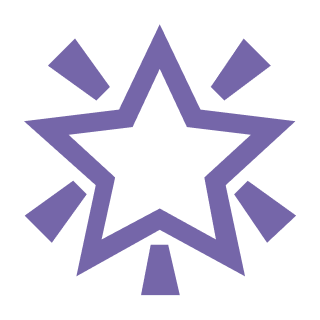}} \textbf{Disappearance}: Disappearance describes the gradual fading of an object from the screen. Objects can transition from being fully visible to transparent, or they can shrink in size gradually. Disappear animations are often used to conclude a scene or conceal unnecessary objects.

\subsubsection{{Dimension II: Action Classification}} The storyline involves a variety of intricate and diverse actions, making it impractical to delineate the design space for each specific action. Hence, our aim is to conceptualize specific actions for generalization. For example, ``\textit{Tom \uline{runs} from home to school}'' and ``\textit{Tom \uline{walks} from home to school}'' represent distinct specific actions. However, both actions fall within the broader concept of ``\textit{changing the location of an object}''. This methodology, known as Conceptual Dependency Theory (CDT)~\cite{schank1972conceptual}, was introduced by RC Schank and has been widely adopted in numerous studies~\cite{cambria2014jumping,taylor2022schematic,graesser1994constructing}. The theory outlines $11$ conceptual primitives for actions. Considering the expression capabilities of motion comics, we combined several similar concepts and ultimately identified $8$ fundamental types of actions.

\begin{itemize}
    \item \textbf{ATRANS (Abstract TRANSformation)}: to change an abstract relationship of an object.
    \item \textbf{PTRANS (Position TRANSformation)}: to change the location of an object.
    \item \textbf{PROPEL}: to apply a force to an object.
    \item \textbf{MOVE}: to move a body part.
    \item \textbf{INGEST}: to take something inside an object.
    \item \textbf{EXPEL}: to take something from inside an object and force it out.
    \item \textbf{SPEAK}: to produce a sound.
    \item \textbf{MENTAL}: to transfer information mentally or to create or combine thoughts.
\end{itemize}

\par After establishing the analysis dimensions, we initially identified $1031$ specific actions from the aforementioned $95$ motion comics. Subsequently, we categorized each specific action according to its fundamental type and recorded the atomic animation operations utilized. Finally, we compiled the frequency of combinations of fundamental types and atomic operations, resulting in the analysis findings presented in \autoref{tab:design_result}.

\begin{table*}[h]
  \centering
  \caption{Statistical frequency of combinations of atomic animation operations and fundamental action types.}
  \label{tab:design_result}
  \resizebox{\textwidth}{!}{
  \begin{tabular}{|c|c|c|c|c|c|c|c|}
    \hline
    \diagbox[width=15em]{\textbf{Action Type}}{\textbf{Atomic Operation}} & \textbf{Path Movement} & \textbf{Scale} & \textbf{Rotation} & \textbf{Flip} & \textbf{Appearance} & \textbf{Disappearance} & \textbf{\uline{Total}} \\
    \hline
    \textbf{ATRANS} & $24$ & $0$ & $0$ & $0$ & $6$ & $4$ & $34$\\
    \hline
    \textbf{PTRANS} & $593$ & $24$ & $0$ & $0$ & $99$ & $18$ & $734$\\
    \hline
    \textbf{PROPEL} & $112$ & $0$ & $80$ & $0$ & $32$ & $0$ & $224$\\
    \hline
    \textbf{MOVE} & $106$ & $32$ & $395$ & $18$ & $0$ & $0$ & $551$\\
    \hline
    \textbf{INGEST} & $20$ & $0$ & $0$ & $0$ & $0$ & $32$ & $52$\\
    \hline
    \textbf{EXPEL} & $15$ & $0$ & $0$ & $0$ & $46$ & $0$ & $61$\\
    \hline
    \textbf{SPEAK} & $8$ & $23$ & $0$ & $0$ & $168$ & $0$ & $199$\\
    \hline
    \textbf{MENTAL} & $6$ & $8$ & $0$ & $0$ & $76$ & $0$ & $90$\\
    \hline
    \textbf{\uline{Total}} & $884$ & $87$ & $475$ & $18$ & $427$ & $54$ & $1945$\\
    \hline
  \end{tabular}
  }
\end{table*}

\begin{figure}[h]
    \includegraphics[width=\linewidth]{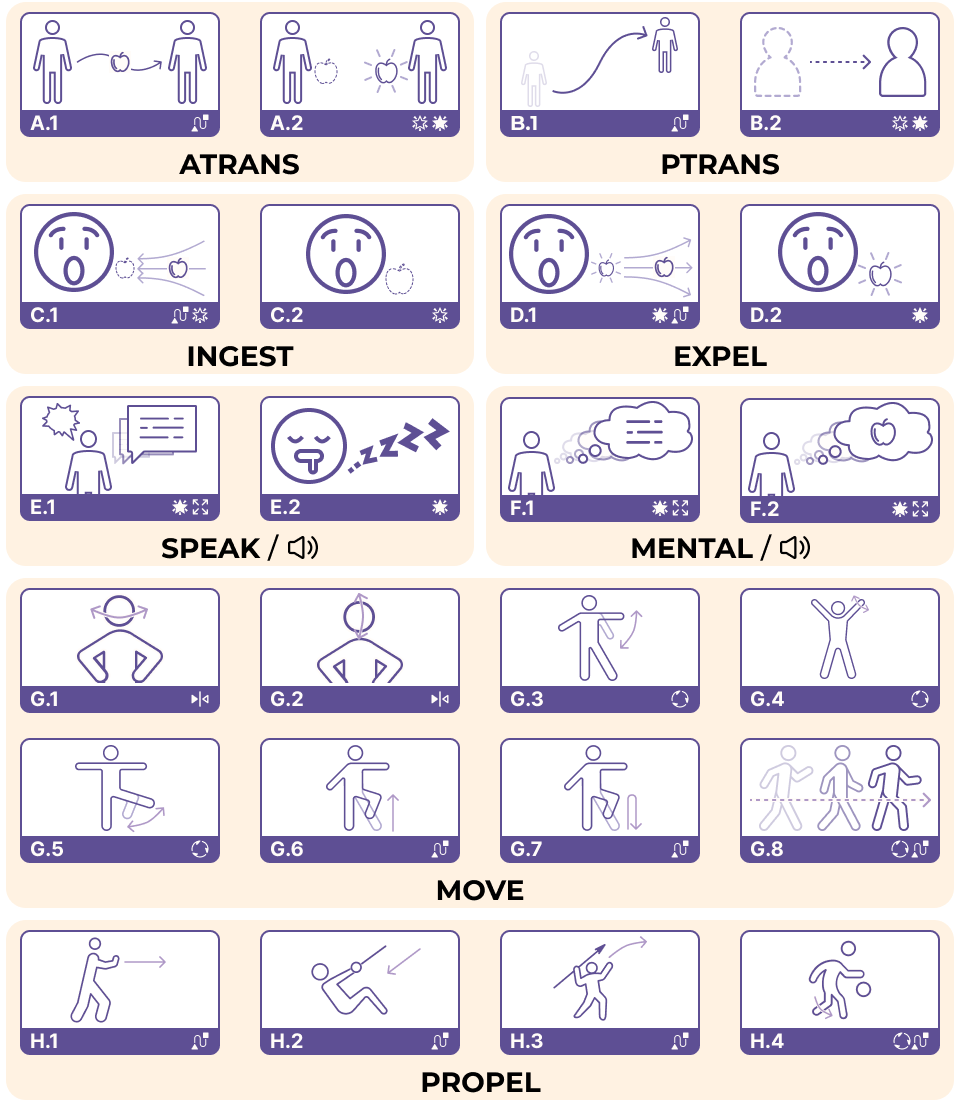}
    \caption[]{The design space for object actions is organized into categories based on six atomic animations and eight fundamental types of action. Each card displays icons in the bottom-right corner indicating the atomic animations that constitute the action.}
    \label{fig:design_space}
\end{figure}

\subsection{Design Space for Object Actions}

\par Based on the results and insights obtained during the analysis process, we collaborated with two experts (\textbf{E3--4}) to pinpoint design patterns associated with each fundamental type of action.

\subsubsection{{The pattern of \uline{ATRANS} and \uline{PTRANS} are similar.}}
\par The design patterns of \textbf{ATRANS} can be summarized into two categories: (1) illustrating the movement of an object moving from the sender to the recipient using path animation \raisebox{-0.7ex}{\includegraphics[height=3ex]{figs/path.png}} (\autoref{fig:design_space} (A.1)); (2) depicting the transition by causing the object to vanish \raisebox{-0.7ex}{\includegraphics[height=3ex]{figs/disappear.png}} from the sender's side and reappear \raisebox{-0.7ex}{\includegraphics[height=3ex]{figs/appear.png}} next to the recipient (\autoref{fig:design_space} (A.2)).

\par Similarly, the design pattern of \textbf{PTRANS} can also be summarized into two categories: (1) showing the movement of an object from its initial position to a new position along a path \raisebox{-0.7ex}{\includegraphics[height=3ex]{figs/path.png}} (\autoref{fig:design_space} (B.1)); (2) making an object disappear \raisebox{-0.7ex}{\includegraphics[height=3ex]{figs/disappear.png}} from the initial position and then reappear \raisebox{-0.7ex}{\includegraphics[height=3ex]{figs/appear.png}} at a new position (\autoref{fig:design_space} (B.2)).

\subsubsection{{\uline{INGEST} and \uline{EXPEL} have reversible patterns.}}
\par \textbf{INGEST} is typically depicted as a combined process involving ``path movement \raisebox{-0.7ex}{\includegraphics[height=3ex]{figs/path.png}} + disappearance \raisebox{-0.7ex}{\includegraphics[height=3ex]{figs/disappear.png}}''. In the scenario of ``\texttt{A [INGEST] B}'', it is depicted on the screen as B moving from its original position along a specific path towards A, followed by disappearance (\autoref{fig:design_space} (C.1)). Alternatively, if the distance between them is relatively close, B may directly vanish (\autoref{fig:design_space} (C.2)).

\par \textbf{EXPEL} serves as the inverse of \textbf{INGEST} in terms of design pattern, typically depicted as a sequence of ``appearance \raisebox{-0.7ex}{\includegraphics[height=3ex]{figs/appear.png}} + path movement \raisebox{-0.7ex}{\includegraphics[height=3ex]{figs/path.png}}'' (\autoref{fig:design_space} (D.1)).

\subsubsection{{\uline{SPEAK} and \uline{MENTAL} rely on specific visual embellishments.}}
\par \textbf{SPEAK} typically incorporates dialogue boxes and textual elements. These dialogue boxes may either materialize \raisebox{-0.7ex}{\includegraphics[height=3ex]{figs/appear.png}} directly or with a scaling transition \raisebox{-0.7ex}{\includegraphics[height=3ex]{figs/scale.png}}, with variations in style depending on the intended tone and atmosphere (\autoref{fig:design_space} (E.1)). Moreover, when conveying onomatopoeic speech, the option exists for text to manifest directly, without the use of a dialogue box (\autoref{fig:design_space} (E.2)).

\par In the design space of \textbf{MENTAL}, the thoughts are consistently encapsulated within bubbles, which can manifest \raisebox{-0.7ex}{\includegraphics[height=3ex]{figs/appear.png}} directly or with a scaling transition \raisebox{-0.7ex}{\includegraphics[height=3ex]{figs/scale.png}}. These thought bubbles may contain either simple text (\autoref{fig:design_space} (F.1)) or visual elements (\autoref{fig:design_space} (F.2)).

\par It's also worth noting that some creators prefer to directly present the dialogue, onomatopoeic or thoughts directly with audio (\raisebox{-0.7ex}{\includegraphics[height=3ex]{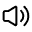}}).

\subsubsection{{\uline{MOVE} and \uline{PROPEL} have extremely complex design spaces.}}
\par The design patterns associated with \textbf{MOVE} and \textbf{PROPEL} are notably intricate, often requiring the repetition and combination of atomic animation operations. For example, the action of striding, which falls under \textbf{MOVE}, involves a sequence involving lifting, advancing, and alternating steps between left and right feet, incorporating diverse combinations of movement along intricate paths. \textbf{E3} commented, ``\textit{For amateurs, implementing such complex animations is extremely difficult}''. Following discussions with \textbf{E3} and \textbf{E4}, we decided to distill common patterns from the intricate design space and integrate corresponding animation operations for each pattern (\autoref{fig:design_space} (G), \autoref{fig:design_space} (H)).

\section{\textit{DancingBoard}}
\par We propose \textit{DancingBoard}, an integrated authoring system that provides a multi-stage pipeline to facilitate the creation of motion comics for amateur creators. Based on the previous obtained design goals, \textit{DancingBoard} adopts a concise interface design (\textbf{DG1}), follows a typical workflow (\textbf{DG2}) and gives users sufficient guidance and assistance (\textbf{DG3, DG4}) at each step to help reduce their workflow and cognitive costs.

\subsection{System Overview}
\par The \textit{DancingBoard} system consists of two pages: the \textit{Upload Page} (\autoref{fig:overview} (A)) and the \textit{Creation Page} (\autoref{fig:overview} (B)). The \textit{Upload Page} allows users to upload story content in two ways: either by typing directly into the input box or by uploading documents in \texttt{.txt} format. Once the story is uploaded, the main functions for creation are provided by the \textit{Creation Page}, which consists of three views: (1) \textbf{Outline Browser} (\autoref{fig:overview} (B.1)), which represents the story content and outlines by scene; (2) \textbf{Visual Element Editor} (\autoref{fig:overview} (B.2)), which provides previews of visual elements (including characters and items) in the story, and supports editing; and (3) \textbf{Canvas Workspace} (\autoref{fig:overview} (B.3)), which displays the content of motion comics and supports various editing operations.

\subsection{Outline Browser}
\label{sec:outline}
\par The Outline Browser (\autoref{fig:overview} (B.1)) provides a comprehensive overview of the story's original text, including key information and its outline. Using GPT-4o, the essential data is extracted from the story text, by segmenting the narrative text of the story into scenes and presenting them in the form of cards. If users find any discrepancies in the segmentation results provided by GPT-4o, they have the option to click the \raisebox{-1ex}{\includegraphics[height=3ex]{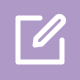}} button to make necessary modifications. To facilitate quick retrieval of key information from lengthy text, the system employs various visual cues to distinguish \colorbox{characterColor}{characters}, \colorbox{itemColor}{items}, \colorbox{actionColor}{actions}, and \colorbox{dialogueColor}{dialogue \& thoughts} with each scene.

\begin{figure}[h]
    \includegraphics[width=\linewidth]{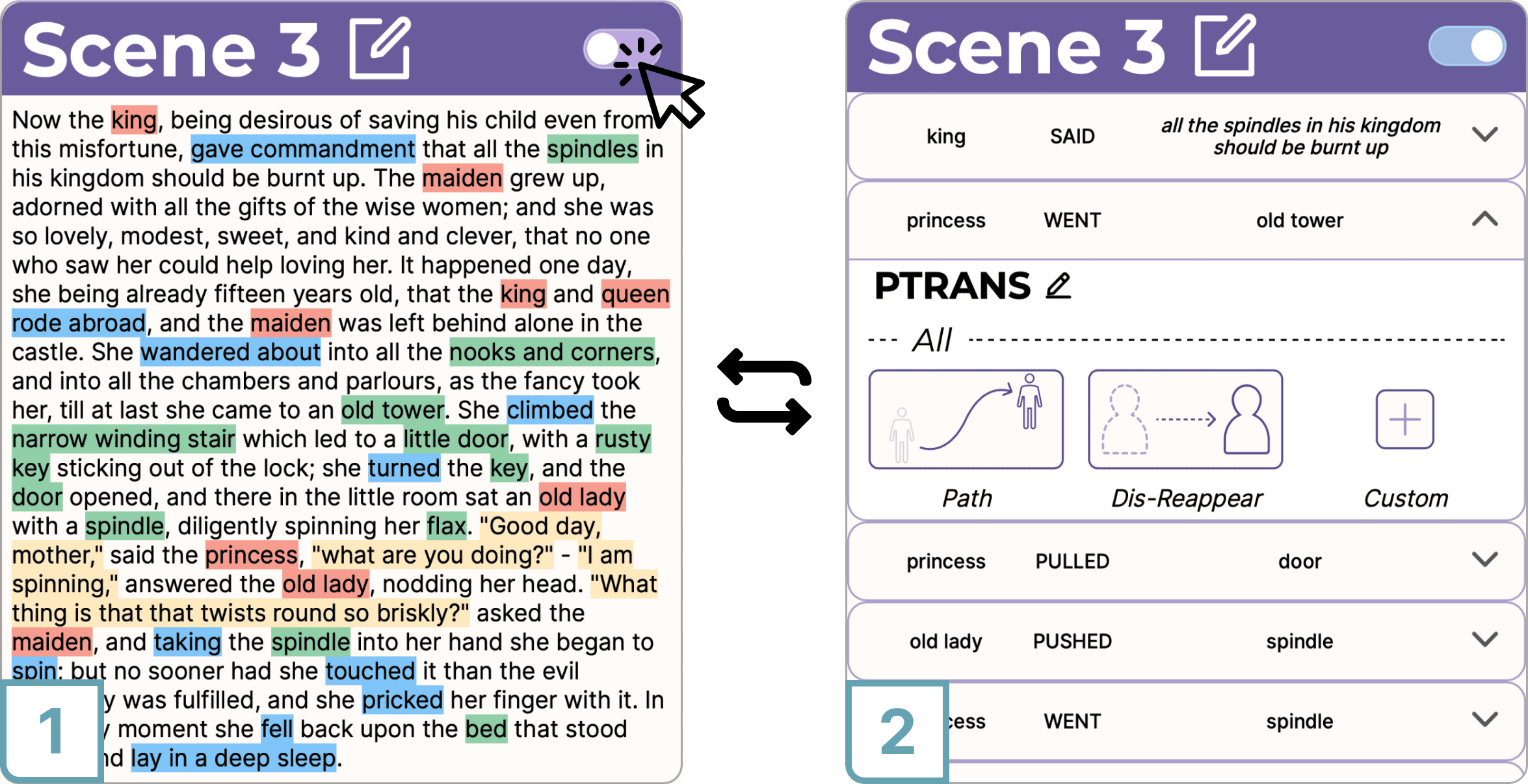}
    \caption[]{The Outline Browser facilitates seamless toggling between two modes: (1) displaying the original text and (2) presenting an outline for each scene. In the outline view (2), all specific actions are listed chronologically from top to bottom. Users can access built-in animation suggestions by clicking the \raisebox{-1ex}{\includegraphics[height=3ex]{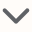}} button.}
    \label{fig:outline_detail}
\end{figure}

\par Furthermore, leveraging the extracted actions along with the actors and recipients, the system provides a concise outline representation for each story segment within each scene. As depicted in \autoref{fig:outline_detail} (1), users have the flexibility to switch between the original text and the outline by clicking the \raisebox{-1ex}{\includegraphics[height=3ex]{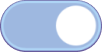}} button. As shown in \autoref{fig:outline_detail} (2), specific actions within each scene are arranged chronologically from top to bottom, aiding users in structuring a coherent narrative from extensive text content, thus facilitating  planning for motion comics. For each action listed in the outline, users can view built-in animation suggestions by clicking the \raisebox{-1ex}{\includegraphics[height=3ex]{figs/Buttons/Dropdown_Button.png}} button. For a specific action, the system employs GPT-4o to categorize each action into one of the eight fundamental types and subsequently provides relevant animation suggestions based on the design space (\autoref{fig:design_space}). For example, the specific action ``\textit{princess WENT old tower}'' in \autoref{fig:outline_detail} (2) falls under the \textbf{PTRANS} category, prompting the system to propose two animation suggestions: \textit{Path} and \textit{Dis-Reappear}. Should users find the classification results unsatisfactory, they can modify the category by clicking the \raisebox{-1ex}{\includegraphics[height=3ex]{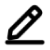}} button. These suggestions streamline the expression of character actions within the Canvas Workspace, with further usage details elaborated in \autoref{sec:workspace}.

\subsection{Visual Element Editor}
\par The Visual Element Editor (\autoref{fig:overview} (B.2)) facilitates the editing and previewing of visual elements within the motion comics. It's structured into two rows: the upper row showcases character previews, while the lower row displays item previews.

\par
\begin{figure}[h]
    \includegraphics[width=\linewidth]{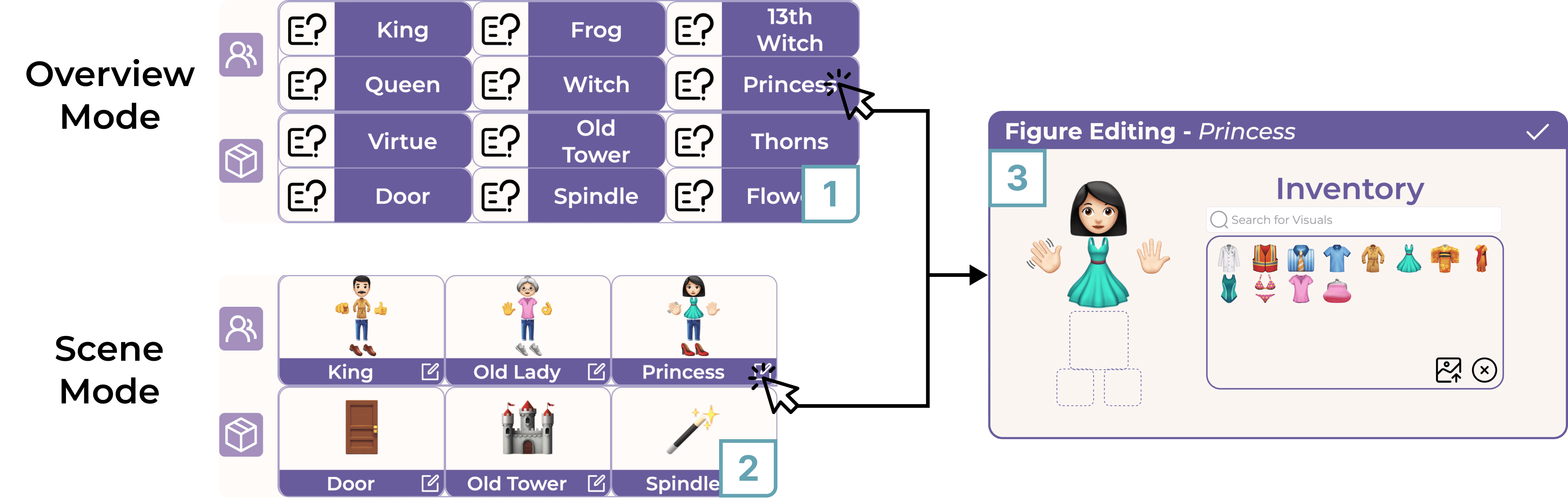}
    \caption[]{The Visual Element Editor offers two distinct modes: (1) Overview Mode, which presents all characters and items in the story when no scene is selected, and (2) Scene Mode, which exclusively showcases visual elements pertinent to the selected scene. In either mode, users can seamlessly transition to the editing panel by clicking, enabling them to adjust the visual design of elements.}
    \label{fig:editor_detail}
\end{figure}

 When no scene is selected in the Outline Browser, the Visual Element Editor defaults to overview mode (\autoref{fig:editor_detail} (1)), displaying all characters and items in the story. Users can modify their visual designs by clicking on their respective names. \clf{As shown in \autoref{fig:editor_detail} (3), \textit{DancingBoard} includes a built-in visual asset library for users to choose from. In this paper, we use emojis as a representative example of visual assets due to their wide coverage and expressive versatility~\cite{bai2019systematic, kralj2015sentiment, erle2022emojis, RIORDAN201775}. Additionally, users can choose other visual assets or upload their custom assets via the \raisebox{-1ex}{\includegraphics[height=3ex]{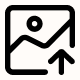}} button.} We advise users to consider the mobility of different body parts when crafting character prototypes to enhance the portrayal of character actions, while keeping the skeleton simple for amateur creators to manipulate. Thus, the system facilitates users in individually adjusting the head, body, and limbs of characters to create prototypes with enhanced mobility. Users can also customize a single part, such as using only a head. It is worth noting that for the same visual element, users can design multiple prototypes to accommodate varying scene requirements. When a scene in the Outline Browser is selected, the Visual Element Editor transitions to scene mode (\autoref{fig:editor_detail} (2)), displaying only the visual elements relevant to the corresponding scene. These elements can be easily dragged and added to the Canvas Workspace.

\subsection{Canvas Workspace}
\label{sec:workspace}

\par
\begin{figure*}[h]
    \includegraphics[width=\linewidth]{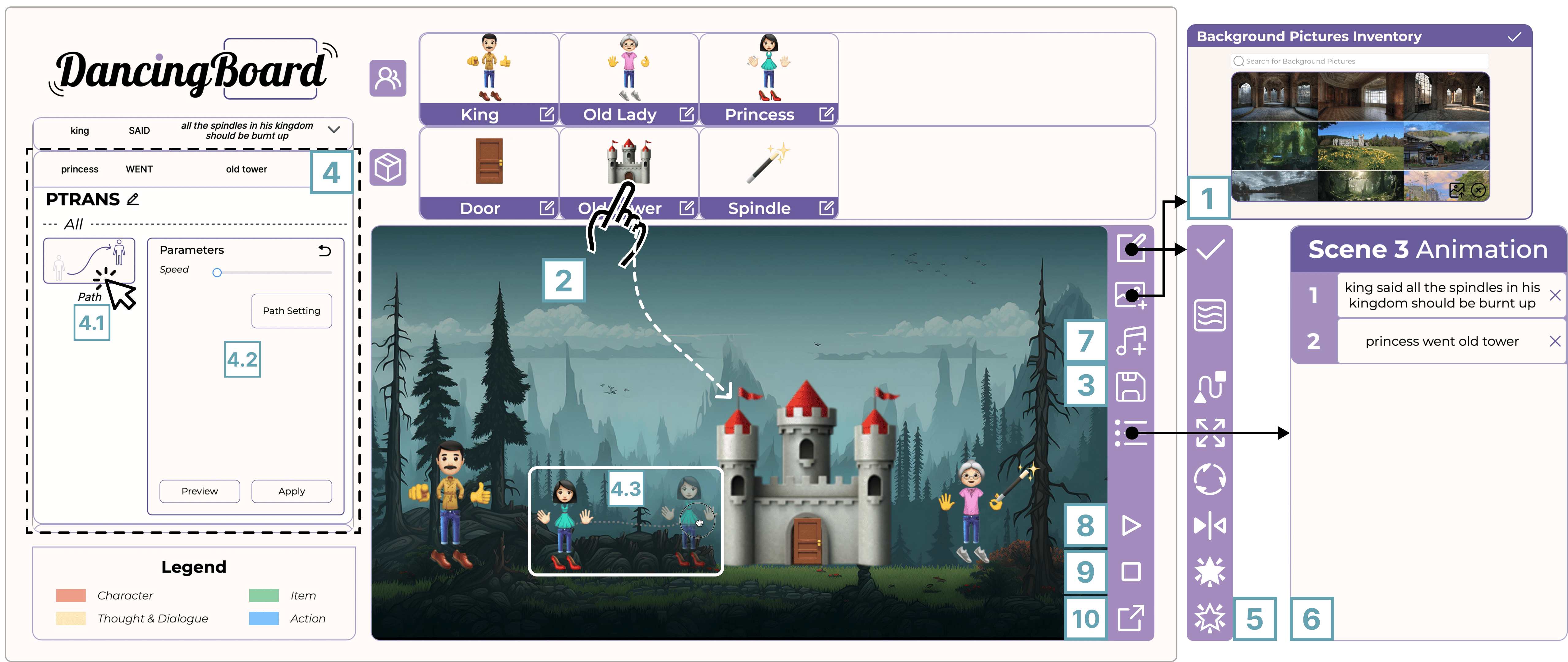}
    \caption[]{The Canvas Workspace provides essential features for creating motion comics, including: (1) Accessing the image asset library to choose suitable background images. (2) Dragging visual elements to appropriate positions within the scene. (3) Saving the basic layout of visual elements for easy retrieval. (4) Easily applying animation effects to visual elements based on suggestions. (5) Performing atomic animation operations on elements seamlessly. (6) Expanding or collapsing the animation list, which outlines the sequence of animations. (7) Uploading the background music (BGM) for the scene. (8) Initiating playback of all animations in the scene from the beginning. (9) Halting animation playback and reverting the canvas to its original state. (10) Exporting the content of the current scene of motion comic for further use.}
    \label{fig:workspace_detail}
\end{figure*}

 The Canvas Workspace (\autoref{fig:overview} (B.3)) provides displaying, editing, and adjustment for each scene, including background image, BGM, layout of visual elements, and animation effects. Upon selecting a scene segment in the Outline Browser, the Canvas Workspace dynamically presents the relevant content. Initially, the workspace appears blank, allowing users to choose appropriate background images by clicking the \raisebox{-1ex}{\includegraphics[height=3ex]{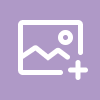}} button (\autoref{fig:workspace_detail} (1)). Subsequently, users can easily drag visual elements from the Visual Element Editor and position them as desired within the workspace (\autoref{fig:workspace_detail} (2)).

\par Once users have finalized the basic layout of visual elements and saved their configuration by clicking the \raisebox{-1ex}{\includegraphics[height=3ex]{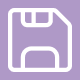}} button (\autoref{fig:workspace_detail} (3)), they can proceed to add animation effects. As outlined in \autoref{sec:outline}, animation suggestions are available for each action in the Outline Browser. As shown in \autoref{fig:workspace_detail} (4), taking the action ``\textit{princess WENT old tower}'' as an example, users can access the animation control interface by selecting one of the action options (\autoref{fig:outline_detail} (2)). \textit{``WENT''} falls into the category of \textbf{PTRANS}, which has two options: \textit{Path} and \textit{Dis-Reappear}.

\begin{figure}[h]
    \centering
    \includegraphics[width=\linewidth]{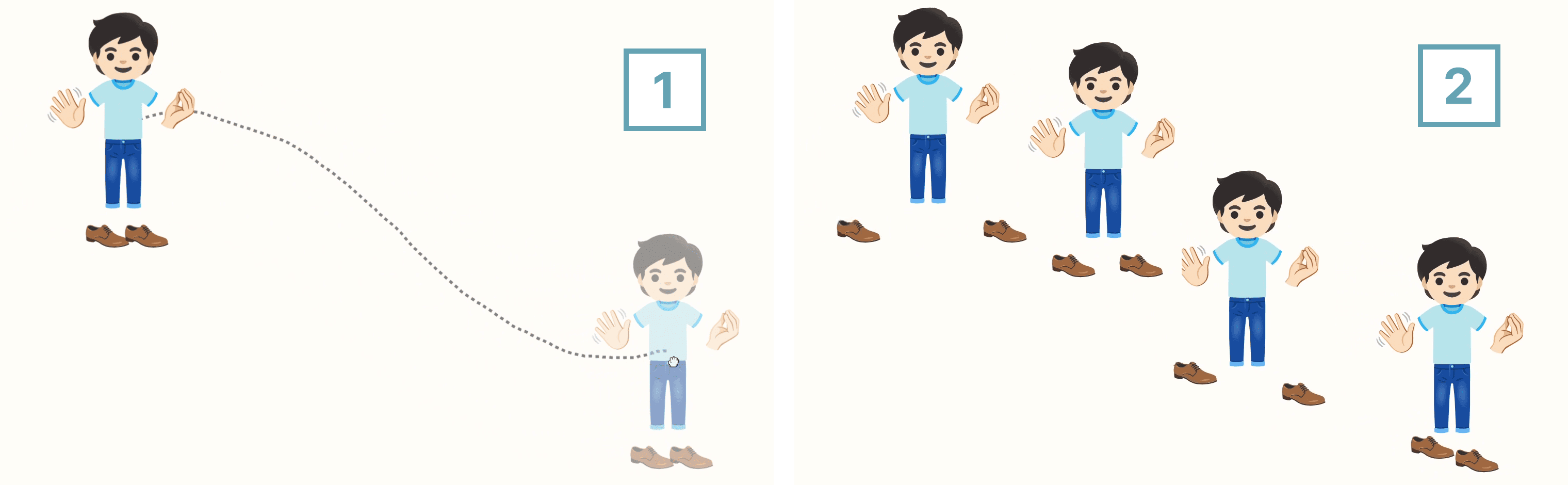}
    \caption[]{The animation of \textit{Path}. (1) The user specifies the action’s trajectory by dragging the element, which is visually represented by a dotted line. (2) During the animation, the character alternates its feet to simulate walking along the defined path.}
    \label{fig:animation}
\end{figure}

\par As an example, consider the animation for \textit{Path} \raisebox{-1.3ex}{\includegraphics[height=4ex]{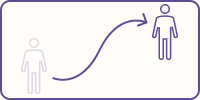}} (\autoref{fig:workspace_detail} (4.1)). This corresponds to the Design Space discussed in \autoref{sec:designspace}. When \textit{Path} is applied to an element, the element moves from its current position to the destination, following a trajectory specified by the user, which is indicated by a dotted line during editing(\autoref{fig:animation} (1)). Furthermore, if \textit{Path} is applied to a character whose lower limbs are implemented, an animation of alternating feet simulates walking along the path (\autoref{fig:animation} (2)).

\par Within the outline browser (\autoref{fig:workspace_detail} (4)), users can control the animation with the parameters provided (\autoref{fig:workspace_detail} (4.2)). The `speed' parameter regulates the velocity of the princess's movement, while the `path setting' parameter allows user to determine the trajectory along which the princess travels (\autoref{fig:workspace_detail} (4.3)). After configuring the parameters, users can preview the animation effect in the workspace by clicking the \raisebox{-1ex}{\includegraphics[height=3ex]{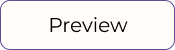}} button. Once satisfied with the preview, users can apply the animation effect to the relevant elements by clicking the \raisebox{-1ex}{\includegraphics[height=3ex]{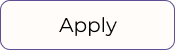}} button. In addition to utilizing animation suggestions, users can also employ animation atomic operations on elements by clicking the \raisebox{-1ex}{\includegraphics[height=3ex]{figs/Buttons/Edit_Button.png}} button located on the right side (\autoref{fig:workspace_detail} (5)).

\par To streamline management, the system provides an animation list (\autoref{fig:workspace_detail} (6)), which catalogs all animation effects applied to elements within the current scene. Sequential numbers on the left denote the order in which these animations occur. Users can promptly remove corresponding animations by clicking \raisebox{-0.3ex}{\includegraphics[height=2ex]{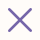}}. To conserve space, the animation list is not constantly visible. Users can toggle its display by clicking the \raisebox{-1ex}{\includegraphics[height=3ex]{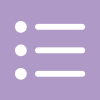}} button located on the right side of the workspace. Additionally, the \raisebox{-1ex}{\includegraphics[height=3ex]{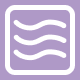}} button enables users applying artificial blur for creating spatial depth, and \raisebox{-1ex}{\includegraphics[height=3ex]{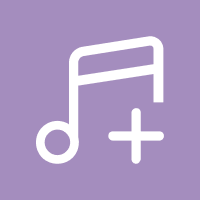}} button allows users to upload the background music for the scene. When one edition of the scene is finished, the \raisebox{-1ex}{\includegraphics[height=3ex]{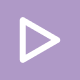}} button on the right side allows users to initiate playback of all animations in the list from the beginning (\autoref{fig:workspace_detail} (8)), while the \raisebox{-1ex}{\includegraphics[height=3ex]{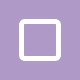}} button halts playback and restores the canvas to its initial state (\autoref{fig:workspace_detail} (9)). Finally, the \raisebox{-1ex}{\includegraphics[height=3ex]{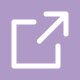}} button facilitates the export of the content of the current scene content (\autoref{fig:workspace_detail} (10)).

\subsection{Implementing Details}
\par In the back-end engine, we employed the \textit{Flask}\footnote{https://flask.palletsprojects.com/en/3.0.x/} framework in conjunction with the GPT-4o API to process user-input or uploaded story text. We deconstructed the analysis task into three sequential steps: (1) Segmenting the story text into distinct scenes based on shifts in the narrative environment; (2) Extracting key information from these segmented scenes, including characters, items, actions, dialogues, and thoughts; (3) Categorizing the extracted actions into eight fundamental action types we've defined. To streamline task completion, we meticulously devised prompts adhering to OpenAI's prompt engineering principles~\cite{gptpromptengineer}. \autoref{app:A} provides a detailed list of prompt phrases associated with each task.

\par In the front-end system, we developed the web-based \textit{DancingBoard} system utilizing the responsive \textit{Vue.js}\footnote{https://vuejs.org/} framework. For interactive animations, we leveraged the \textit{GreenSock Animation Platform}\footnote{https://gsap.com/} (\textit{GSAP}), a JavaScript animation library renowned for its ability to deliver high-performance animations in web browsers. \textit{DancingBoard} also supports text-to-speech for characters voicing using \textit{Azure Speech Services}\footnote{https://learn.microsoft.com/en-us/azure/ai-services/speech-service/get-started-text-to-speech}.

\begin{figure*}[h]
    \includegraphics[width=\linewidth]{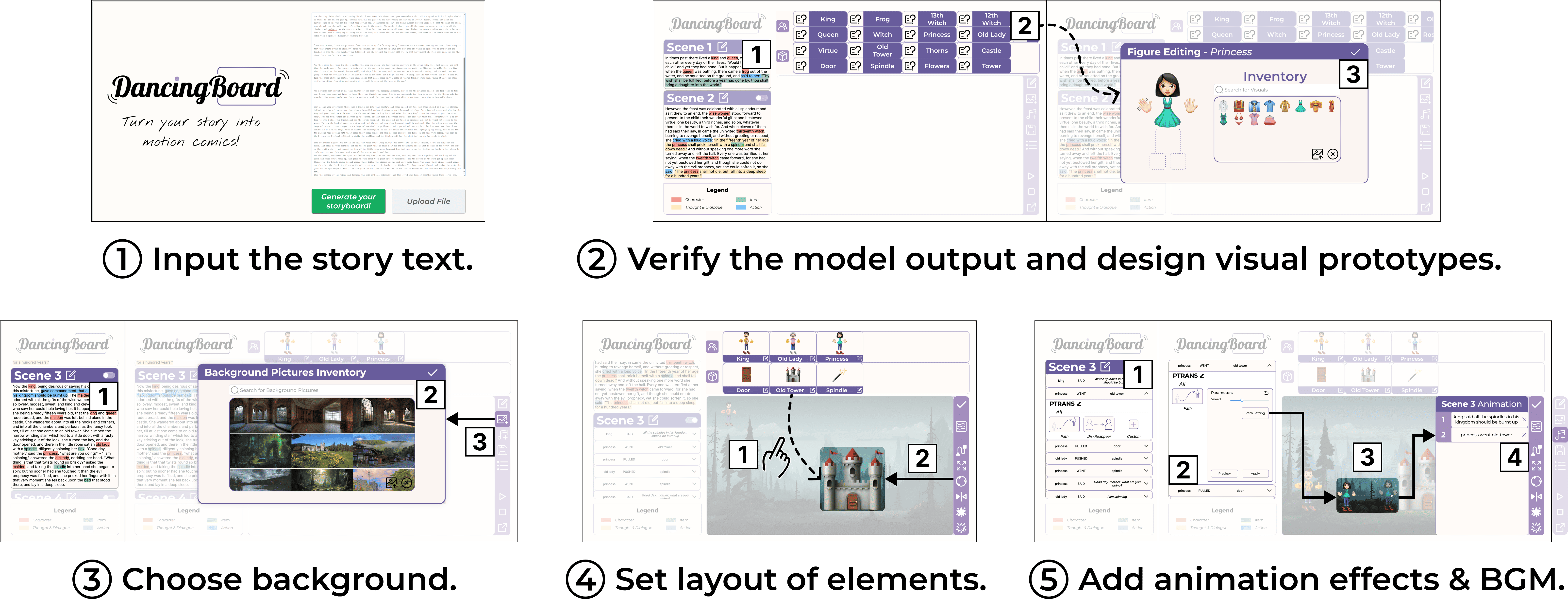}
    \caption[]{The procedure for a user to craft motion comics with \textit{DancingBoard} is sequentially outlined starting with \textcircled{1}, where the narrative text is inputted. At \textcircled{2}, the user assesses the precision of key information extracted by the LLM before proceeding to edit the depictions of characters and items. Moving to \textcircled{3}, users can then select each scene to customize its background. The adjustment of elements’ placement and properties is conducted at \textcircled{4}. The final step, \textcircled{5}, involves adding and adjusting animations, thereby finalizing the motion comic.}
  \label{fig:walkthrough}
\end{figure*}

\subsection{System Walkthrough}
\par We illustrate the utilization of \textit{DancingBoard} by walking through a typical usage scenario, as shown in \autoref{fig:walkthrough}: Consider Daniel, an amateur short video creator with limited experience in crafting motion comics. When he first learned about motion comics, he was attracted, and wanted to transform the existing stories (here \textit{Sleeping Beauty}) into motion comics for a short video. However, with little prior knowledge on the creation process, he is uncertain about where to start. Consequently, Daniel turns to \textit{DancingBoard} for assistance. 

\par Starting with the source material, Daniel inputs the story text into the \textit{Upload Page} (\autoref{fig:walkthrough} (1)). After a brief processing period, the system transitions to the \textit{Creation Page}, presenting the extracted key information in the Outline Browser and Visual Elements Editor. Daniel first verifies the contents in the Outline Browser to ensure that the system accurately segmented the scene and detected the key information (\autoref{fig:walkthrough} (\textcircled{2}.1)). Subsequently, he proceeds to the Visual Element Editor, where he can edit the visual design of each entity (\autoref{fig:walkthrough} (\textcircled{2}.2)). As shown in \autoref{fig:walkthrough} (\textcircled{2}.3), for each empty slot, Daniel browses through the library of assets to select the most appropriate options. For example, he decides a skirt asset for the Body slot to represent the princess. Following this process, Daniel creates prototypes for each entity in the story, which will be utilized in subsequent creations.

\par After completing the visual prototype, Daniel chooses a specific scene (\autoref{fig:walkthrough} (\textcircled{3}.1)) to animate. He begins by choosing a suitable background image from the resource library for this scene (\autoref{fig:walkthrough} (\textcircled{3}.2)). Subsequently, he arranges all characters and items within the Canvas Workspace (\autoref{fig:walkthrough} (\textcircled{4}.1)), adjusting their positions and organizing a preliminary layout based on the narrative of the scene (\autoref{fig:walkthrough} (\textcircled{4}.2)), and clicked \raisebox{-1ex} {\includegraphics[height=3ex]{figs/Buttons/Save_Scene_Button.png}} button to save the scene after finalizing the layout. Activating the \raisebox{-1ex}{\includegraphics[height=3ex]{figs/Buttons/Switch_Button.png}} button (\autoref{fig:walkthrough} (\textcircled{5}.1)), Daniel toggles to the outline representations. An outline displaying ``\textit{princess WENT old tower}'' is expanded to depict this action and present the built-in animation options. The term \textit{WENT} belongs to the fundamental type \textbf{PTRANS}. According to our proposed design space, it corresponds to two common animation narratives (\autoref{fig:design_space} (B.1) \& (B.2)). Daniel selects the `path' option \raisebox{-1.3ex}{\includegraphics[height=4ex]{figs/Buttons/ptrans_all_path.png}} (\autoref{fig:walkthrough} (\textcircled{5}.2)) to visualize this action. To avoid a simplistic straight-line trajectory, he employs the `Path Setting' feature \raisebox{-1.3ex}{\includegraphics[height=4ex]{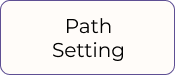}}, enabling him to define the movement by dragging the princess towards the old tower (\autoref{fig:walkthrough} (\textcircled{5}.3)), where the pathway is recorded during this operation. After configuring the path, he modifies the `speed' parameter to regulate the motion speed. Lastly, he previews the action using the \raisebox{-1ex} {\includegraphics[height=3ex]{figs/Buttons/Preview_Button.png}} button, where he sees the princess walk towards the old tower step by step along the pathway. Satisfied with the outcome, he applies the animation to the scene by clicking the \raisebox{-1ex}{\includegraphics[height=3ex]{figs/Buttons/Apply_Button.png}} button, thereby saving this animation to the list (\autoref{fig:walkthrough} (\textcircled{5}.4)).

\par Daniel repeats these steps for the remaining outline representations within this scene: selecting built-in animations, adjusting parameters, previewing, and saving them. Despite a few actions being incorrectly classified into alternative fundamental types, he easily corrects them. For actions that cannot be directly depicted through predefined animations, Daniel manually constructs them using the atomic operations provided by the system (\autoref{fig:walkthrough} (\textcircled{5}.4)). After finalizing the edits for all pending actions, he chooses a BGM that fit this scene best with the \raisebox{-1ex} {\includegraphics[height=3ex]{figs/Buttons/BGM.png}} button, and saves the scene. Upon completing the editing of all scenes, Daniel initiates the playback of the animations in sequence by clicking the \raisebox{-1ex} {\includegraphics[height=3ex]{figs/Buttons/Play_Button.png}} button. Watching the animations run smoothly in the Workspace, Daniel feels satisfied and relieved.

\section{User Study I: Evaluation of system effectiveness in the creative process}
\par To answer \textbf{(RQ4) How effectively does \textit{DancingBoard} support the creation of motion comics?}, we conducted a user study under controlled conditions, evaluating and comparing participants' creative experiences using \textit{DancingBoard} against those using a baseline system.

\subsection{Participants}
\par We recruited $23$ amateur creators (denoted as \textbf{C1--23}, comprising $7$ females and $16$ males) from the participants of the formative study. Their ages spanned from $18$ to $27$ years, with an average of $22.2$ years ($SD=3.02$). All respondents possessed some experience in creating motion comics, but none of them could be considered proficient. The detailed population could be seen at \autoref{tab:Study1Participants} in \autoref{app:User_Study1}.

\subsection{Baseline Condition}

\par \clf{After reviewing the authoring tools in \autoref{sec:2.1}, we sought a baseline system that strikes a balance between ease of use for amateurs and the expressive power of motion comics. On the one hand, professional tools like \textit{Photoshop} and \textit{After Effects} offer extensive functionalities but demand significant manual effort. On the other hand, AIGC text-to-video tools enable ``one-click generation'' but provide limited editing capabilities, especially for fine-tuning visual elements. We also considered intermediate options such as \textit{PowerPoint}, which offers built-in animation effects primarily for slide transitions but lacks the expressiveness needed for dynamic comics. Similarly, \textit{Figma} requires substantial effort to produce smooth animations. In comparison, \textit{Cartoon Animator} combines professional-grade features with user-friendly functionalities such as skeletal system and camera control, achieving an optimal balance between simplicity and creative flexibility. This makes it the ideal baseline for our user study.}

\subsection{Procedure}
\par Initially, after obtaining consent, participants viewed a pre-recorded $30$-minute instructional video detailing the usage of \textit{DancingBoard} and \textit{Cartoon Animator}. They were then allotted $30$ minutes to explore and become accustomed to the functionalities of both systems, with encouragement to seek clarification or ask questions as needed. Following this phase, participants were tasked with completing two separate assignments using each system individually.

\begin{itemize}
    \item \textbf{Task 1:} Participants were prompted to freely explore and utilize the systems, with the goal of creating a motion comic using the provided story content. They were allowed to incorporate visual assets from any available source to facilitate their creations.
    \item \textbf{Task 2:} This task simulated scenarios requiring modifications to the design of visual elements within the content. Following \textbf{Task 1}, participants were tasked with altering the design of specific visual elements in the motion comic while maintaining the integrity of the content expression.
\end{itemize}

\par The detailed task descriptions can be found in \autoref{app:User_Study1}. Participants were instructed to use one system to complete both tasks before switching to the other system. To counteract potential order effects, half of the participants were directed to begin with \textit{DancingBoard} followed by \textit{Cartoon Animator}, while the remaining participants followed the reverse order. We recorded the time taken by participants to complete each task. Upon finishing all tasks, participants were asked to complete a post-task questionnaire, where they rated various aspects of usability, learnability, user experience, and workload of both systems using a five-point Likert scale. The question statements, as shown in \autoref{tab:questions}, were inspired by System Usability Scale (SUS)~\cite{brooke1996sus} and NASA Task Load Index (NASA-TLX)~\cite{hart1986nasa}. To gain deeper insights, we conducted semi-structured interviews with participants. In the end, each participant was compensated with $\$10$.

\par All of the aforementioned procedures were conducted online. Participants accessed \textit{DancingBoard} through web browsers and accessed \textit{Cartoon Animator} through clients. With their consent, the entire user study was recorded and saved as screen and audio recordings.

\begin{table*}[h]
\centering
\caption{Question statements in the post-task questionnaire.}
\label{tab:questions}
\begin{tabular}{c|l}
\hline
\multirow{4}{*}{\textbf{Usability}} & Q1: I thought the system was easy to use.\\ 
& Q2: I found the various functions in this system were well integrated. \\ 
& Q3: I thought the system improved the efficiency of my work.\\ 
& Q4: I thought the system improved the quality of my work.\\ \hline
\multirow{3}{*}{\textbf{Learnability}} & Q5: I needed to learn a lot of things before I could get going with this system.\\ 
& Q6: I think that I would need the support of a technical person to be able to use this system.\\ 
& Q7: I would imagine that most people would learn to use this system very quickly. \\ \hline
\multirow{4}{*}{\textbf{User Experience}} & Q8: I think that I would like to use this system frequently. \\ 
& Q9: I found the system unnecessarily complex.\\ 
& Q10: I thought there was too much inconsistency in this system.\\ 
& Q11: I felt very confident using the system.\\ \hline
\multirow{6}{*}{\textbf{Workload}} & Q12: How mentally demanding was the task? \\ 
& Q13: How physically demanding was the task? \\ 
& Q14: How successful were you in accomplishing what you were asked to do? \\
& Q15: How hard did you have to work to accomplish your level of performance? \\ 
& Q16: How insecure, discouraged, irritated, stressed, and annoyed were you? \\ \hline
\end{tabular}
\end{table*}

\subsection{Results and Analysis}

\par
\begin{figure}[h]
    \includegraphics[width=\linewidth]{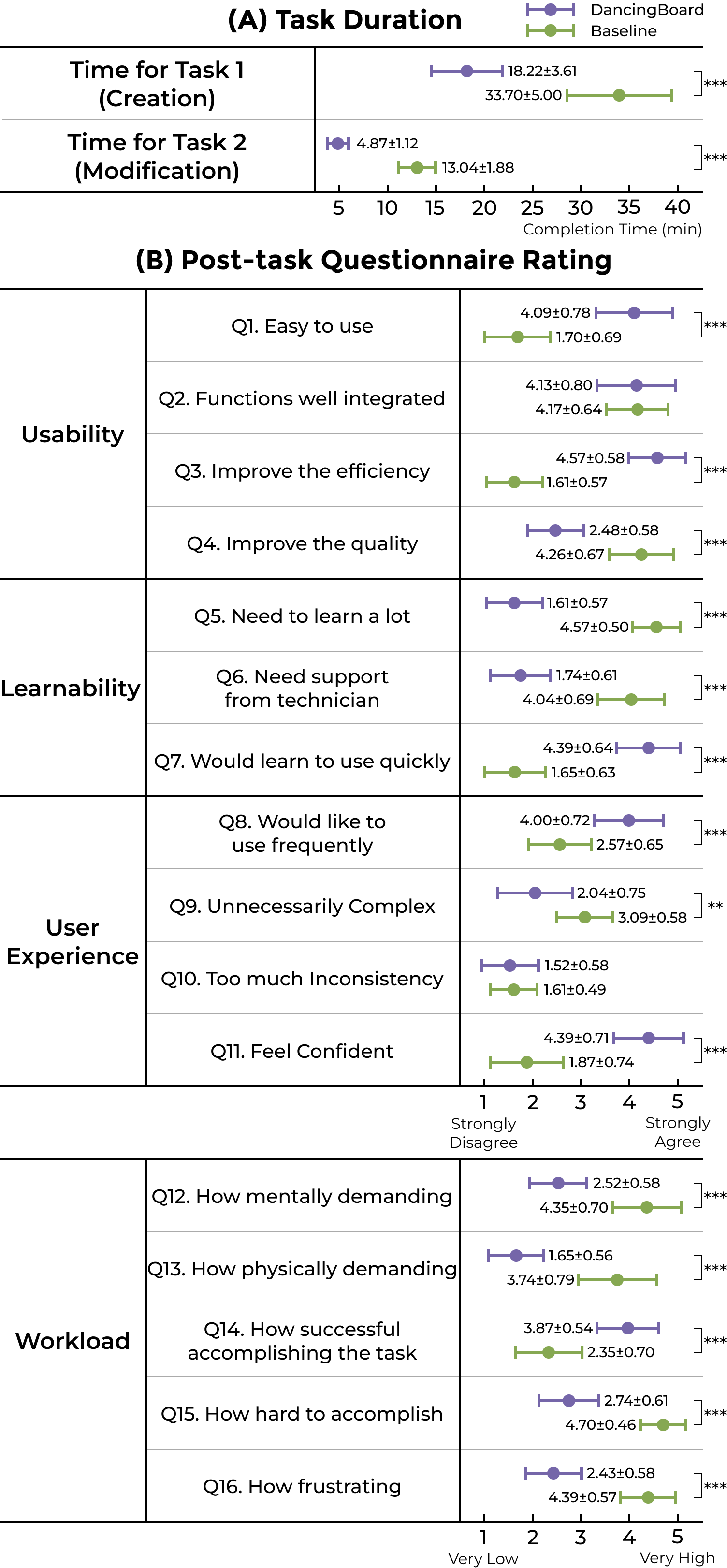}
    \caption[]{The results of User Study I includes two main aspects: (A) the time participants spent completing the tasks using the two systems separately, and (B) the ratings of the post-task questionnaire. ($**: p < .05, ***: p < .01$)}
    \label{fig:user_study_result}
\end{figure}

 We gathered data on the time spent by each participant on each task, along with the scores from the post-task questionnaires. For these data, we performed Wilcoxon matched-pairs signed ranks test, with normality check by Shapiro-Wilk test. Based on the analysis results shown in \autoref{fig:user_study_result} and insights from the semi-structured interviews, we drew the following conclusions.

\subsubsection{Usability and Learnability}
\par In terms of the \uline{average time} spent on \textbf{Task 1} and \textbf{Task 2}, \textit{DancingBoard} demonstrated a significant advantage (\textbf{Task 1}: $Mean=18.22, SD=3.61$ (min); \textbf{Task 2}: $Mean=4.87, SD=1.12$ (min)), far exceeding the baseline system (\textbf{Task 1}: $Mean=33.70, SD=5.00$ (min); \textbf{Task 2}: $Mean=13.04, SD=1.88$ (min)). Based on our observations, for \textbf{Task 1}, most users spent a considerable amount of time on illustrating the animation effect when using the baseline system. In contrast, \textit{DancingBoard}'s pre-defined animation helped users reduce the workload of these steps. The built-in material libraries also saved time for users on searching visual assets.  \textbf{C2} commented, ``\textit{Those animations were really helpful. They not only saved me a lot of time but also clearly presented the story.}'' Regarding the modification of visual designs for characters (\textbf{Task 2}), \textit{DancingBoard} also provided better support. \textbf{C9} mentioned, ``\textit{In Cartoon Animator, it is hard to transfer the animation from one object to another. Therefore, I have to re-add the animation effects. But in DancingBoard, I can simply finish the task by modifying in the visual element editor.}''

\par The results of the post-task questionnaire indicate that \textit{DancingBoard} received wide recognition from participants in terms of \uline{usability} (Q1) (\textit{DancingBoard}: $Mean=4.09, SD=0.78$; Baseline: $Mean=1.70, SD=0.69$). The majority of participants believed that \textit{DancingBoard} could enhance the \uline{efficiency} (Q3) of their creation (\textit{DancingBoard}: $Mean=4.57, SD=0.58$; Baseline: $Mean=1.61, SD=0.57$). \textbf{C2}, \textbf{C7} and \textbf{C11} particularly praised our proposed animation design space based on character actions, stating that it can represent ``\textit{most character actions encountered in the task}''. There was no significant difference observed on the \uline{integration} (Q2) (\textit{DancingBoard}: $Mean=4.13, SD=0.80$; Baseline: $Mean=4.17, SD=0.64$), and users expressed that \textit{Cartoon Animator} could produce better outcomes in terms of \uline{quality} (Q4) when compared to \textit{DancingBoard} (\textit{DancingBoard}: $Mean=2.48, SD=0.58$; Baseline: $Mean=4.26, SD=0.67$). \textbf{C6} noted, ``\textit{Although creating motion comics with professional tools like Cartoon Animator would be challenging for amateurs, but for professional creators, those complex functionalities was the only choice for them to produce high-level works.}''

\par In terms of the learning curve, \textit{DancingBoard} demonstrates its user-friendliness for amateurs. After watching the instructional video, All participants are able to \uline{quickly} (Q7) grasp our system with less \uline{learning efforts} (Q5), without the need for \uline{additional guidance} (Q6). (Q5: \textit{DancingBoard}: $Mean=1.61, SD=0.57$; Baseline: $Mean=4.57, SD=0.50$; Q6: \textit{DancingBoard}: $Mean=1.74, SD=0.61$; Baseline: $Mean=4.04, SD=0.69$; Q7: \textit{DancingBoard}: $Mean=4.39, SD=0.64$; Baseline: $Mean=1.65, SD=0.63$.)

\subsubsection{User Experience}
\par We evaluated user experience based on satisfaction (Q8), system complexity (Q9), inconsistency (Q10), and emotional experience (Q11). In terms of \uline{satisfaction}, participants showed a preference for using \textit{DancingBoard} frequently to create motion comics over \textit{Cartoon Animator} (\textit{DancingBoard}: $Mean=4.00, SD=0.72$; Baseline: $Mean=2.57, SD=0.65$). Participants mentioned ``\textit{convenient}'' and ``\textit{ease of use}'' as positive aspects of our system.  \textbf{C3} said, ``\textit{Although both DancingBoard and Cartoon Animator are tailor-made tools for motion comics, as a new-comer, I would favour tools that are easy to get started with.}'' Regarding \uline{system complexity}, almost no participants considered \textit{DancingBoard} to be an unnecessarily complex system (\textit{DancingBoard}: $Mean=2.04, SD=0.75$; Baseline: $Mean=3.09, SD=0.58$). \textbf{C10} highly praised the interface design, ``\textit{The interface of DancingBoard is very clear and straightforward, hardly causing me any cognitive barriers.}'' In terms of \uline{emotional experience}, the majority of participants provided positive feedback for \textit{DancingBoard} (\textit{DancingBoard}: $Mean=4.39, SD=0.71$; Baseline: $Mean=1.87, SD=0.74$). \textbf{C20} pointed out, ``\textit{DancingBoard provides me with a clear workflow, as if there's someone guiding me step by step on the screen.}'' However, some participants expressed concerns about certain features of the system. For example, \textbf{C5} said, ``\textit{I'm not sure what the LLM will bring. If it provides an inaccurate outline and users don't discover the issue promptly, it will introduce significant uncertainty to the creative process.}'' Most participants agreed that there is little \uline{inconsistency} in both \textit{DancingBoard} and \textit{Cartoon Animator}. (\textit{DancingBoard}: $Mean=1.52, SD=0.58$; Baseline: $Mean=1.61, SD=0.49$.)

\subsubsection{Cognitive Workload}
\par The comparison of cognitive load between the two systems in \autoref{fig:user_study_result} reveals significant statistical differences across dimensions of \uline{mental demand} (Q12) (\textit{DancingBoard}: $Mean=2.52, SD=0.58$; Baseline: $Mean=4.35, SD=0.70$), \uline{physical demand} (Q13) (\textit{DancingBoard}: $Mean=1.65, SD=0.56$; Baseline: $Mean=3.74, SD=0.79$), \uline{own performance} (Q14) (\textit{DancingBoard}: $Mean=3.87, SD=0.54$; Baseline: $Mean=2.35, SD=0.70$), and \uline{effort} (Q15) (\textit{DancingBoard}: $Mean=2.74, SD=0.61$; Baseline: $Mean=4.70, SD=0.46$). Participants widely acknowledge that \textit{DancingBoard} integrates practical features for creating motion comics, contributing to faster task completion. \textbf{C2} noted a significant reduction in the workload associated with retrieving visual elements and designing animation effects while using \textit{DancingBoard}. \textbf{C23} found the Outline Browser highly practical, commenting, ``\textit{It is like a to-do list, guiding me through each step.}'' \textbf{C7} and \textbf{C8} also expressed that \textit{DancingBoard} helped them maintain a good workflow pace. \uline{Performance} (Q16) of \textit{DancingBoard} remained superior to the baseline system (\textit{DancingBoard}: $Mean=2.43, SD=0.58$; Baseline: $Mean=4.39, SD=0.57$). \textbf{C9} expressed that using \textit{Cartoon Animator} to complete \textbf{Task 2} was ``\textit{doing endless repetitive work}'' which ``\textit{drove him nuts}''.

\subsubsection{Suggestions for \textit{DancingBoard}}
\par In the semi-structured interviews, participants also pointed out some areas for improvement in \textit{DancingBoard} to make the process of creating motion comics more adaptable and user-friendly.

\par \textbf{Enhance narrative timeline.} Currently, \textit{DancingBoard} only presents animations in a serial list format within each scene. Users can adjust their order by dragging and preview them by clicking. However, some participants find it challenging to perceive the specific duration of each animation. \textbf{C3} suggested adding specific duration to the right of each animation in the list. Meanwhile, some participants proposed ideas on supporting multiple animation effects concurrently. \textbf{C12} noted, ``\textit{Not just in motion comics, some cool visual effects are made by combining multiple animations. A timeline integrating these, like in Adobe Premiere Pro, would be helpful.}''

\par \textbf{Consider cinematographic narration.} The current system only considers enhancing narration by adding animation effects to visual elements. However, many participants mentioned that cinematographic language also played a significant role in some motion comics they had seen before, especially in emphasizing and transitioning. For instance, and placing a person in the center of the screen and zooming in on him speaking  to emphasize their importance. Therefore, they suggested that \textit{DancingBoard} could introduce some simple camera effects, such as zooming in, zooming out, and tracking, which could improve the visual presentation and narrative effects. Such improvements would allow users to more flexibly control camera movements, making the outcome more vivid and engaging.

\section{User Study II: Delivery evaluation of created content}

\par To answer \textbf{(RQ5) Can the motion comics created using \textit{DancingBoard} accurately convey stories to the audience?}, we conducted another user study under controlled conditions, evaluating and comparing the audiences' narrative comprehension with the motion comics created by \textit{DancingBoard} and by professional tools.

\subsection{Participants}
\par We recruited $18$ audience members (denoted as \textbf{A1--18}, comprising $10$ females and $8$ males) \clf{via a partner university's email list}. Their ages spanned from $18$ to $24$ years, with an average of $20.5$ years ($SD=1.64$). \clf{None of the participants were acquaintances of the authors, and they had limited prior knowledge of motion comics.}

\subsection{Baseline}
\par We selected two professional motion comics from \textit{YouTube}, respectively \textit{The Accountant}~\cite{Accountant} (denoted as \texttt{Pro-Accountant}) and \textit{A Night Out With SPIDER-MAN \& WOLVERINE}~\cite{ANightOutSpiderman} (denoted as \texttt{Pro-Spiderman}). Before conducting further experiments, we ensured that no participants had either viewed the motion comics or gained any knowledge directly related to the works, such as the original comic from which the motion comics were adapted. This was done to avoid any potential impacts effects of prior knowledge.

\par
\begin{figure}[h]
    \centering
    \includegraphics[width=\linewidth]{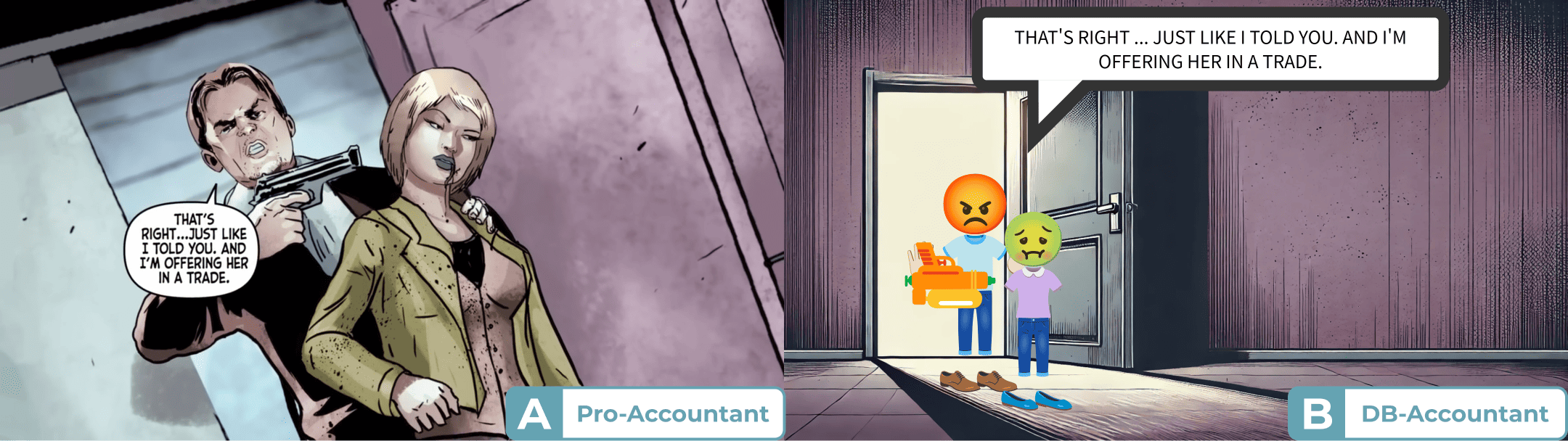}
    \caption[]{An example scene of (A) \texttt{Pro-Accountant} and (B) \texttt{DB-Accountant} which narrated the same story.}
    \label{fig:study2example}
\end{figure}

 One author reproduced the both works as separate motion comics (denoted as \texttt{DB-Accountant} and \texttt{DB-Spiderman}) with \textit{DancingBoard} which narrated the exact same stories. \autoref{fig:study2example} is an example scene of (A) \texttt{Pro-Accountant} and (B) \texttt{DB-Accountant}.

\subsection{Procedure}
\par The study began with an introduction to the concept of motion comics. Participants were then equally divided into two groups. The first group was assigned to watch \texttt{Pro-Accountant} first, followed by \texttt{DB-Spiderman}, while the second group watched \texttt{DB-Accountant} first and then \texttt{Pro-Spiderman}. This arrangement was made to minimize potential order effects. After viewing each motion comic, participants were required to complete a questionnaire. The design of the questionnaire was inspired by previous research~\cite{song2022visual, kim2017visualizing, qiang2017storytelling} and investigated feedback on motion comics across three sections:

\begin{itemize}
    \item \textbf{Comprehension}: This section evaluated how well participants understood the story conveyed by the motion comic. Participants were asked to summarize the story as fully as possible based on the following narrative elements: \uline{settings} (space \& time), \uline{characters}, \uline{core conflict}, and \uline{events}.

    \item \textbf{Retention}: This section evaluated the efficiency with which participants could retrieve information from the story. Participants were asked to complete six single-choice questions about specific details of the story. Both the \uline{answer accuracy} and \uline{completion time} were recorded. The questions can be found in the \autoref{app:User_Study2}.

    \item \textbf{Reaction}: This section gauged participants' subjective reactions to the motion comic. Participants were asked to rate three five-point Likert scale questions including (1) \uline{interest}: (how curious to know more about the comic ($1 - bored, 5 - interested$); (2) \uline{hedonic}: how satisfied about the comic ($1 - unpleasant, 5 - Enjoyable$); (3) \uline{satisfaction}: how satisfied about the comic ($1 - dissatisfied, 5 - satisfied$).
\end{itemize}

\par After completing all tasks, each participant underwent a semi-structured interview to gather further feedback.

\par The entire procedures were conducted via \textit{Zoom}, with the motion comic materials distributed to participants through \textit{OneDrive}. It is important to note that participants were not allowed to watch the same motion comic multiple times, nor were they allowed to fill out the questionnaire while viewing the comic. With the participants' consent, the entire user study process was recorded, and both screen and audio recordings were saved. As compensation, each participant received a \$5 reward.

\subsection{Results and Analysis}
\par
\begin{figure}[h]
    \includegraphics[width=\linewidth]{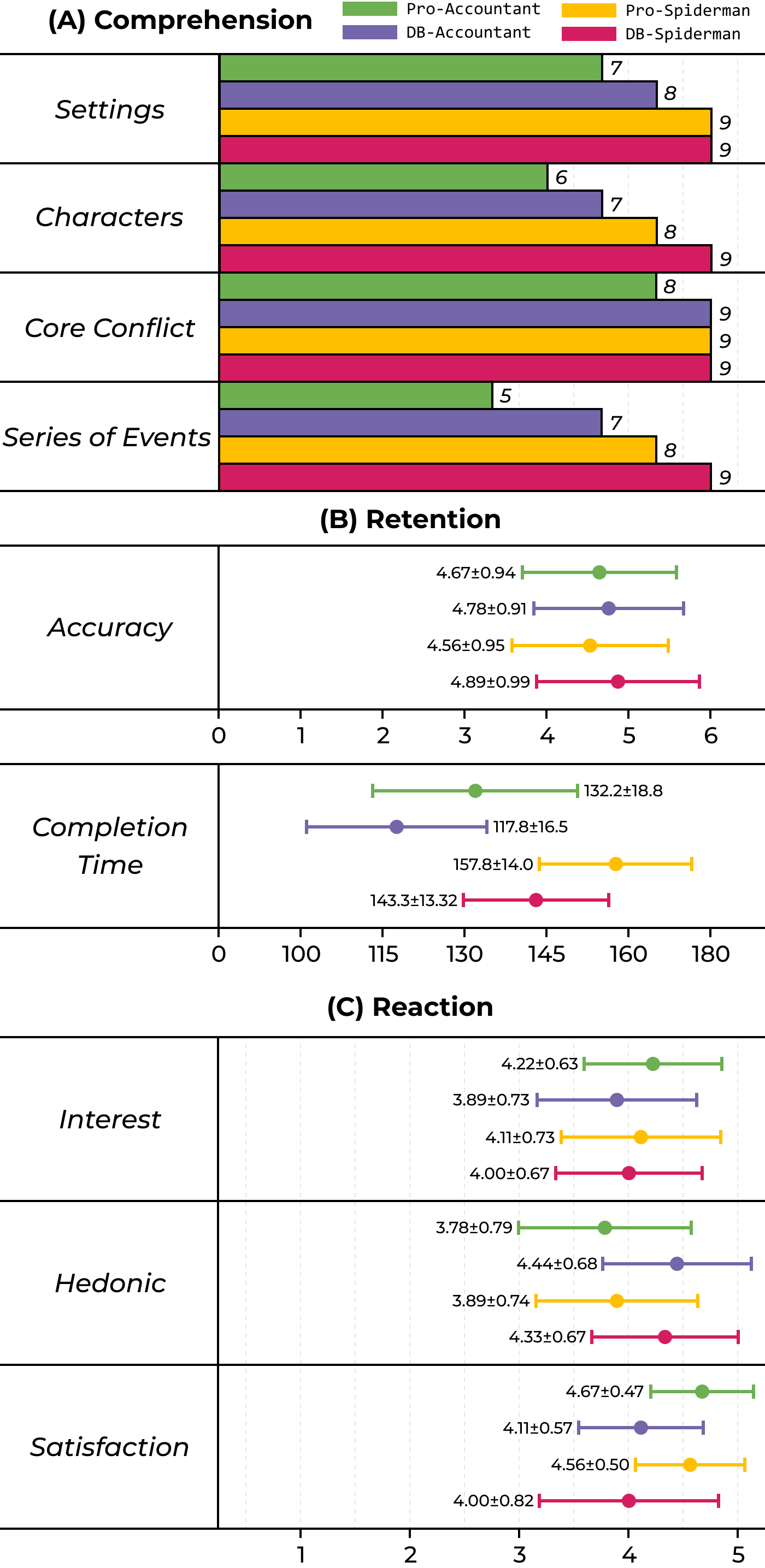}
    \caption[]{The results of User Study II includes three main aspects: (A) Comprehension, (B) Retention, and (C) Reaction. No significant difference observed.}
    \label{fig:user_study2_result}
\end{figure}

 For the collected data, we performed Wilcoxon matched-pairs signed ranks test, with normality check by Shapiro-Wilk test. Based on the analysis results shown in \autoref{fig:user_study2_result} and insights from the semi-structured interviews, we confirmed that the resulting motion comics by \textit{DancingBoard} were generally as effective as the professional ones in communicating the story to the audience.

\par Furthermore, we drew the following conclusions:

\subsubsection{Comprehension}
\par Results indicate that the majority of participants successfully identified the narrative elements in both motion comics produced by \textit{DancingBoard} and those created with professional tools. For the \uline{settings}, among the 9 participants in each group, 7 recognized the settings in \texttt{Pro-Accountant}, 8 identified those in \texttt{DB-Accountant}, while both \texttt{Pro-Spiderman} and \texttt{DB-Spiderman} had all 9 participants correctly identifying their settings. Regarding \uline{characters}, 6 participants recognized those in \texttt{Pro-Accountant}, 7 identified those in \texttt{DB-Accountant}, 8 recognized characters in \texttt{Pro-Spiderman}, and all 9 participants identified characters in \texttt{DB-Spiderman}. When it comes to the \uline{core conflict}, 8 participants identified it in \texttt{Pro-Accountant}, and all 9 participants recognized it in both \texttt{DB-Accountant} and \texttt{Pro-Spiderman}, with \texttt{DB-Spiderman} also receiving 9 identifications. Finally, for the \uline{events}, 5 identified those in \texttt{Pro-Accountant}, 7 recognized events in \texttt{DB-Accountant}, while both \texttt{Pro-Spiderman} and \texttt{DB-Spiderman} had 8 identifications each. Overall, the results suggest no significant difference between the two versions of the same story.

\par Notably, most participants agreed that the visual simplicity of \textit{DancingBoard}'s outcome helped them grasp the story more quickly. \textbf{A4} commented, ``\textit{You see, the first one (\texttt{Pro-Accountant}) was somehow fancy. To create a tense atmosphere, some visual effects were added to the narrative, which made me distracted. In contrast, the simplicity of later one (\texttt{DB-Spiderman}) can let me catch these elements directly.}'' Some participants noted that the original story also affected the comprehension. \textbf{A10} said, ``\textit{The story of Spiderman is long, and Accountant has lots of scenarios involved. Both of them posed challenge for me.}''

\subsubsection{Retention}
\par In terms of \uline{answer accuracy}, results show that participants could catch the details of the story from both versions of motion comics, where no significant difference observed between them (\texttt{Pro-Accountant}: $Mean=4.67, SD=0.94$, \texttt{DB-Accountant}: $Mean=4.78, SD=0.91$, \texttt{Pro-Spiderman}: $Mean=4.56, SD=0.95$, \texttt{DB-Spiderman}: $Mean=4.89, SD=0.99$). Regarding \uline{completion time}, results indicated that participants generally finished the tasks faster with \textit{DancingBoard}'s outcome than with professional ones (\texttt{Pro-Accountant}: $Mean=132.2, SD=18.76$ (s), \texttt{DB-Accountant}: $Mean=117.8, SD=16.50$ (s), \texttt{Pro-Spiderman}: $Mean=157.8, SD=13.96$ (s), \texttt{DB-Spiderman}: $Mean=143.3, SD=13.32$ (s)). The visual simplicity of \textit{DancingBoard}'s outcome also contributed to this difference. \textbf{A16} mentioned, ``\textit{The professional one (\texttt{Pro-Spiderman}) often conveys information in a subtle way, which needs my further thoughts to understand what happened. However, the simple one (\texttt{DB-Accountant}) directly presented the key messages. For example, taking the painting as a gift for the accountant is demonstrated by the simple action of passing it along the path.}''

\subsubsection{Reaction}
\par In terms of \uline{interest}, participants demonstrated generally positive attitudes towards all motion comics (\texttt{Pro-Accountant}: $Mean=4.22, SD=0.63$, \texttt{DB-Accountant}: $Mean=3.89, SD=0.73$, \texttt{Pro-Spiderman}: $Mean=4.11, SD=0.73$, \texttt{DB-Spiderman}: $Mean=4.00, SD=0.67$), with no significant difference observed between the two versions of the same story. Regarding \uline{hedonic}, results show that most participants had a enjoyable viewing experience for all motion comics (\texttt{Pro-Accountant}: $Mean=3.78, SD=0.79$, \texttt{DB-Accountant}: $Mean=4.44, SD=0.68$, \texttt{Pro-Spiderman}: $Mean=3.89, SD=0.74$, \texttt{DB-Spiderman}: $Mean=4.33, SD=0.67$), where the ratings for \textit{DancingBoard}'s outcome is slightly higher than the professional ones. This can be attributed to the simple visual style of \textit{DancingBoard}'s outcome. \textbf{A9} mentioned, ``\textit{Although the original story (The Accountant) is serious and tense, the visual adaption (by \texttt{DB-Accountant}) really introduced a taste of humor, which reduced my stress.}'' \textbf{A13} commented, ``\textit{My 4-year-old brother would definitely love this.}''

\par For \uline{satisfaction}, results indicated that all motion comics satisfied most audiences (\texttt{Pro-Accountant}: $Mean=4.67, SD=0.47$, \texttt{DB-Accountant}: $Mean=4.11, SD=0.57$, \texttt{Pro-Spiderman}: $Mean=4.56, SD=0.50$, \texttt{DB-Spiderman}: $Mean=4.00, SD=0.82$), but the ratings for \textit{DancingBoard}'s outcome is slightly lower than the professional ones. Participants proposed that the ``\textit{Quick and Dirty}'' style of \textit{DancingBoard} led to unpromising performance. \textbf{A2} noted, ``\textit{(\texttt{DB-Spiderman}) can help me better understand the story through those simplistic visual designs, but I still prefer those created by professionals (\texttt{Pro-Accountant}). Professionals can find the perfect point balancing all the considerations.}''

\section{Discussion}

\par In this section, we reflect on the design of \textit{DancingBoard}, discuss potential research opportunities and future visions.

\subsection{Integration of LLM into Authoring Tools}
\label{sec:LLM}
\par Within \textit{DancingBoard}, we leverage \textit{GPT-4o} for three main tasks: segmenting story text into scenes, extracting key narrative elements (such as characters, items, actions, dialogues, and thoughts), and categorizing specific actions into eight fundamental types. While these tasks might seem manageable for creators, we integrate an LLM for two main reasons. First, it alleviates creators from time-consuming, repetitive tasks, enabling them to shift their focus to more creative aspects of content development. Our user study results confirm this benefit, showing that delegating these tasks to LLM reduces physical workload. Second, it aims to provide a typical workflow of the creative process. As noted in our formative study (\autoref{sec:formative_study}), participants often lacked a structured workflow and tended to dive into creation without sufficient planning, which hindered both efficiency and output quality. \textit{DancingBoard} addresses this by subtly guiding users toward a more systematic approach. Rather than relying on textual guidance alone, we designed a tool that actively facilitates the creation process. By leveraging LLM to assist users in initial planning, the system presents the model's outputs in an organized manner, reducing cognitive and physical load and providing users with concrete examples to help them understand and adopt effective workflows.

\par While the current version of \textit{DancingBoard} only scratches the surface of harnessing LLM's understanding capabilities, we have identified numerous avenues for future development efforts. Prior studies, such as those by Xiao et al.~\cite{xiao2024typedance} and Hou et al.~\cite{hou2024c2ideas}, have demonstrated the significant potential of LLM in supporting visual design. However, in \textit{DancingBoard}, visual element creation remains largely dependent on user creativity. A promising direction for further exploration involves leveraging LLMs to generate a diverse array of alternative visual design options, offering creators an expanded set of ideas and inspiration. Additionally, text-to-video models like \textit{Sora}~\cite{Sora}, although come with their own limitations, still illustrate the potential to generate dynamic content. This suggests the feasibility of using LLMs to produce animations that visually depict actions from natural language descriptions or even to automatically generate entire motion comic directly from textual input.

\par While many authoring tools are increasingly integrating LLMs into their workflows -- whether for text editing in platforms like \textit{Word}~\cite{Word} and \textit{Notion}~\cite{Notion}, or for image editing tools such as \textit{Midjourney}~\cite{Midjourney} and \textit{Firefly}~\cite{Firefly} -- we must also recognize the potential downsides of these technologies. A key concern is hallucination~\cite{huang2023survey}, where LLMs generate inaccurate or misleading outputs based on faulty assumptions. Additionally, since these models are trained largely on internet-sourced data, they may inherit biases and discriminatory content, resulting in skewed or unbalanced outputs. While LLMs can offer convenience and inspiration, as highlighted by participants in our user study, their use should be approached with caution to avoid unintended consequences.

\subsection{``Quick and Dirty'' Prototyping}
\par Our objective in crafting \textit{DancingBoard} was to enhance support for amateur creators by integrating the conventional creation process into a comprehensive authoring tool. This approach offers several notable benefits. First, \clf{as proven in User Study I}, it provides a cost-effective solution by offering an intuitive and user-friendly interface, which translates complex creation processes into simple operational steps. This user-centric design lowers the barrier to entry, allowing creators to unleash their creativity without needing professional technical skills or expertise. Second, \textit{DancingBoard} enables creators to swiftly translate their concepts into tangible prototypes \clf{while keeping their work easy to comprehend as shown in User Study II}, benefiting not only amateur creators but also professionals. For example, professionals can use \textit{DancingBoard} to generate initial draft prototypes during the early stages of motion comic development and refine them through subsequent iterations. Additionally, the system alleviates the workload throughout the creation process, from initial planning to later modification. By offering pre-designed templates, elements, and animation effects, \textit{DancingBoard} streamlines the creation of motion comics, saving time and effort while enabling creators to focus more on content generation and artistic expression.

\par While \textit{DancingBoard} excels in rapid and cost-effective prototyping, it may have limitations regarding detail and precision. For projects that require extensive customization and meticulous adjustments, the tool might not offer adequate support. \clf{Such prototyping methods are often referred to as ``Quick and Dirty'' prototyping, a methodology that focuses on creating fast, low-cost prototypes using readily available materials~\cite{HOLLENDER20101278}. This approach has been shown to be effective for brainstorming and fostering creativity~\cite{gomez2019impossible, HOLLENDER20101278}, enhancing understanding~\cite{QuickandDirtyMultimedia}, and conducting evaluations~\cite{brooke1996sus, baillie2010designing}. \textit{DancingBoard} utilizes this idea by providing users a set of example visual designs and pre-defined animation, allowing amateurs to create easily comprehensible motion comics with lower barriers and less required effort. Through User Study I \& II, we have demonstrated \textit{DancingBoard}'s usability and effectiveness, while user's responses also reflected that the simplicity of \textit{DancingBoard}'s outcome helps them grasp the original story more quickly. Meanwhile, detailed editing is also supported by \textit{DancingBoard} such as user-uploaded visual assets and customized character animation with atomic operations, which takes a step towards better alignment with real-life adoption but also comes with higher cognitive load for amateur users.} Hence, when utilizing \textit{DancingBoard}, amateur creators must strike a balance between the convenience of quick prototyping and the fidelity of their work, ensuring that the final output meets their expectations and specifications.

\par Overall, \textit{DancingBoard} fosters a culture of swift exploration and experimentation with prototypes, encouraging creators to stay adaptable and open to refining or discarding ineffective ideas. Looking ahead, we aim to integrate collaborative features into the system, enhancing inclusivity and facilitating a more participatory prototyping process.

\subsection{Limitation and Future Work}
\par Our research has several limitations that warrant consideration. First, while we employed LLM to analyze story content and perform various tasks, it's crucial to note its potential for inaccuracies, as discussed in \autoref{sec:LLM}. This limitation requires users to carefully verify the outputs generated by the system. Second, the design space, based on $95$ motion comics, may not be fully comprehensive. Future research should include a wider array of examples to enrich the existing animation design patterns and broaden the applicability of our tool. \clf{Third, the system has limited support for creating high-fidelity, professional motion comics, which may not fully meet the needs of users with advanced artistic skills or specific aesthetic goals. The design goal of \textit{DancingBoard} is to enable non-professionals, especially those without prior experience in animation or visual design, to easily create motion comics. This focus inherently involves a trade-off between usability and professional functionality. Currently, the system emphasizes simplicity by offering pre-defined visual assets and animation effects, which might constrain users seeking greater flexibility in visual styles, scene perspectives, or detailed customizations. In the future, we aim to explore ways to incorporate richer features, such as integrating image and video generation models, to enhance the flexibility of creating motion comics while maintaining a low cognitive load. Our goal is to strike a better balance between usability and creative freedom, ultimately achieving a system that embodies the principles of ``\textit{low threshold, high ceiling, wide walls}''~\cite{resnick2005design}.} Fourth, the sample size of our user study is limited, indicating a need for further evaluation. Future work could involve recruiting a larger and more diverse participant pool to compare our tool with other creation tools. \clf{Lastly, current user studies focus on system usability and cognitive load but does not specifically evaluate the system's ability to support creativity. Future research could incorporate the creativity support index~\cite{cherry2014quantifying} to assess how effectively \textit{DancingBoard} fosters creative expression and to better capture its impact on users' creativity.}

Third, the system has limited support for creating high-fidelity, professional motion comics. The design goal of \textit{DancingBoard} is to enable non-professionals to easily create motion comics, which inherently involves a trade-off between usability and professional functionality. In the future, we aim to explore ways to provide richer features while maintaining a low cognitive load for users.

\section{Conclusion}
\par This study systematically explores the creative process of motion comics, highlighting the challenges encountered by amateur creators and suggesting avenues for enhancing the creative experience. Our contribution includes the development of \textit{DancingBoard}, an integrated tool for creating motion comics. It boasts a streamlined interface, adheres to a typical workflow, and offers users comprehensive guidance and support at each stage, thereby reducing their workload and cognitive burden. Furthermore, we distilled design patterns of character actions from $95$ motion comic samples and integrated them as built-in suggestions within the \textit{DancingBoard} system. The findings from a user study involving $23$ amateur creators underscore the efficacy of \textit{DancingBoard} in lowering entry barriers and facilitating creation, and results from another user study involving $18$ audience members confirmed \textit{DancingBoard}'s effectiveness in conveying the story.

\begin{acks}
\par We thank anonymous reviewers for their valuable feedback. This work is supported by grants from the National Natural Science Foundation of China (No. 62372298), Shanghai Engineering Research Center of Intelligent Vision and Imaging, Shanghai Frontiers Science Center of Human-centered Artificial Intelligence (ShangHAI), and MoE Key Laboratory of Intelligent Perception and Human-Machine Collaboration (KLIP-HuMaCo).
\end{acks}

\bibliographystyle{ACM-Reference-Format}
\bibliography{sample-base}

\appendix
\onecolumn

\section{Prompt Design}
\label{app:A}
\par Following the principle of \textit{splitting complex tasks into simpler sub-tasks}, we divided the task of understanding story text using GPT-4o into three sequential steps: (1) Segmenting the story text into different scenes based on changes in the narrative environment; (2) Extracting key information from these segmented scenes, including characters, items, actions, dialogues, and thoughts; (3) Categorizing the extracted actions into eight fundamental types of actions we've defined.

\par For these three tasks, we adhere to the following prompt engineering principles proposed by OpenAI~\cite{gptpromptengineer}:

\begin{enumerate}
    \item \textbf{Include details in your query to get more relevant answers.} By explicitly listing the key considerations, steps to complete the task, and output requirements in the task description, we assist the model in understanding and executing the task more accurately. By clearly indicating content changes and transitions to focus on, the model can more precisely identify and segment scenes in the story.
    \item \textbf{Ask the model to adopt a persona.} We explicitly specify the role of the model in the prompt, aiding the model in better understanding the task's context and requirements, thereby providing more accurate and relevant answers. Through persona setting, the model can resonate better with the task's content and demands, enhancing the effectiveness and quality of task execution.
    \item \textbf{Use delimiters to clearly indicate distinct parts of the input.} We employ clear delimiters and headings, such as ``\textit{Task Details}'', ``\textit{Key Considerations}'', and ``\textit{Steps to Complete the Task}'', in the prompt. This aids the model in understanding and organizing the input information more clearly, thereby improving task efficiency and accuracy. Delimiters and headings help the model quickly identify and locate key information, enabling it to process and analyze data more effectively.
    \item \textbf{Specify the steps required to complete a task.} We explicitly list the specific steps required to complete the task in the prompt, such as ``\textit{read and understand the provided list of sentences}'', ``\textit{identify content changes or transitions in the story}'', ``\textit{segment the story scenes based on these changes}'', and ``\textit{create a motion comic using the identified scenes}''. This provides clear operational guidance to the model, aiding in enhancing task execution efficiency and quality. Specifying concrete steps helps the model to systematically handle and analyze data, reducing errors and unnecessary complexity, thereby improving the accuracy and consistency of the output results.
\end{enumerate}

\subsection{Step 1: Segment the Story Text into Different Scenes Based on Environmental Changes}

\par In this phase, we require GPT-4o to segment the story into distinct scenes based on changes in the narrative environment. While GPT-4o can understand and process both medium-length and shorter stories, its maximum output length is limited. This means that GPT-4o cannot directly return text fragments for different scenes if the story is too long. To address this limitation, we instruct GPT-4o to provide the indices of the starting and ending sentences for each scene. This way, even if the story's length surpasses GPT-4o's output constraint, we can still accurately partition the story into different scenes based on the provided indices. 

\par The structure of the prompt at this stage is as follows:

\begin{itemize}
    \item \textbf{Objective:} Clearly convey to GPT-4o the task objectives we aim to achieve.
    \item \textbf{Key Considerations:} Essential factors that require special attention during task execution.
    \item \textbf{Steps to Complete the Task:} Segmenting the task into specific steps aids in generating more precise and high-quality output results. We further refine the task into three distinct steps to assist GPT-4o in systematically understanding and processing the narrative content.
    \item \textbf{Output Requirements:} Define the output format for scene segmentation, including the IDs of the starting and ending sentences for each scene segment, as well as a unique index for each scene.
    \item \textbf{Attention:} Due to the possibility of GPT-4o prematurely terminating and returning incomplete results during task execution, we emphasize the requirement for GPT-4 to return comprehensive scene segmentation information to ensure the integrity of the output.
\end{itemize}

\par The specific prompt for interacting with GPT-4o is as follows:
\begin{lstlisting}
{
    "role": "system", "content": 
     """
    # Task Details
    **Objective**: 
    You are an analyst. Analyze a list of sentences that are parts of a story to identify and 
    categorize the storyline into distinct scenes based on changes in setting or location 
    transitions.

    **Key Considerations**:
    - Pay attention to changes in setting or transitions between locations.
    - Each change in setting or location transition signifies the end of one scene and the beginning 
    of another.
    
    **Steps to Complete the Task**:
    1. Read and understand the list of sentences provided.
    2. Identify changes in setting or location transitions within the narrative.
    3. Divide the story into scenes based on these changes.

    **Output Requirements**:
    - Provide the identified scenes in JSON format.
    - Each scene should be represented as an object with the following key-value pairs:
      - 'id': A unique identifier for each scene (starting from 0, incrementing by 1 for each 
      subsequent scene).
      - 'begin_index': The starting sentence index of the scene within the original sentences array.
      - 'end_index': The ending sentence index of the scene within the original sentences array.

    **Attention**:
    Ensure that the 'end_index' of the last scene corresponds to the index of the last sentence in 
    the original sentences array.

    ----
    # Output Example
    ```
    [
    {"id": 0, "begin_index": 0, "end_index": 5},
    {"id": 1, "begin_index": 6, "end_index": 12},
    ...
    ]
    ```
    """
}
\end{lstlisting}

\subsection{Step 2: Extract Key Information from Each Scene}
\par In this phase, we aim for GPT-4o to accurately extract the core information from each scene, encompassing characters, items, actions, dialogues, and thoughts. For each scene, we will invoke GPT-4o separately for processing.

\par The structure of the prompt at this stage is as follows:
\begin{itemize}
    \item \textbf{Objective:} Clearly articulate the task objectives we intend to achieve with GPT-4o.
    \item \textbf{Tasks:} Explicitly delineate the specific details GPT-4o is required to extract, encompassing characters, items, actions, dialogues, and potential inner thoughts, ensuring a thorough analysis of each scene.
    \item \textbf{Output Requirements:} Precisely stipulate the format requirements for the output. The output should encompass the characters, pertinent objects, and corresponding \texttt{SVO} (Subject, Verb, Object (and Receiver)) action lists for each scene segment, facilitating the generation of a comprehensive and cohesive scene description.
\end{itemize}

\par The specific prompt for interacting with GPT-4o is as follows:

\begin{lstlisting}
{
    "role": "system", "content": 
    """
    #Objective#: 
    
    You are a motion comic creator. You will be provided with a clip from a story. First, you will need to understand the story and clarify the plot and character behavior. Finally, extract information based on the scene text of a story. 

    ----
    #Tasks#

    **Task 1**: Extract the following necessary information:
    
    1.1 **Character**: Identify and list the names of characters appearing in the scene. For example, 
    provide the character names as: `"character": ["grandmother", "Little Red Riding Hood", "mother"]`.
    
    1.2 **Object**: List objects that interact directly with the characters. Exclude background objects. 
    For instance, consider: `"object": ["a cap made of red velvet"]`.
    
    1.3 **SVO**: Extract key actions using the Subject, Verb, Object, and Receiver format. For instance: 
    `"svo": [{"subject": "grandmother", "verb": "gave", "object": "a little cap made of red velvet", 
    "receiver": "Little Red Riding Hood"}]`. If there's no receiver, use an empty string `""`. When a 
    character transfers an item to another character, the `receiver` is mandatory. Multiple sentences 
    from one character should be represented with multiple SVO objects. Example: `"svo": [{"subject": "
    mother", "verb": "said", "object": "Come Little Red Riding Hood.", "receiver": ""}, {"subject": "
    mother", "verb": "said", "object": 
    "Here is a piece of cake and a bottle of wine. Take them to your grandmother.", "receiver": ""}, 
    ...]`.

    ----
    #Output Requirement#
    
    Provide the results in JSON format based on the given example.
    ```
    {
        "character": ["grandmother", "Little Red Riding Hood", "mother", ...],
        "object": ["a cap made of red velvet", ...],
        "svo": [{"subject": "grandmother", "verb": "gave", "object": "a little cap made of red velvet"}, 
        {"subject": "mother", "verb": "said", "object": "Come Little Red Riding Hood. Here is a piece of 
        cake and a bottle of wine. Take them to your grandmother."}, 
        ...]
    }
    ```
    """
}

\end{lstlisting}

\subsection{Step 3: Categorize Actions into Eight Behavioral Categories}
At this step, we require GPT-4o to categorize the actions (or \textit{``verb''}) in each \texttt{SVO} pair into the eight fundamental types of action. We would invoke GPT-4o for processing the \texttt{SVO} pair of each scene seperately. 

\par The structure of the prompt at this stage is as follows:

\begin{itemize}
    \item \textbf{Objective:} Clearly articulate the goals we seek to accomplish with GPT-4o.
    \item \textbf{Key Considerations:} Explicitly describe the input structure, including the specific details regarding the \texttt{SVO} pairs, and elucidate on the eight fundamental action types designated for verb classification.
    \item \textbf{Output Requirements:} Define with precision the structure and content required for the output, emphasizing the necessity to detail the correlation between each "verb" and its assigned category from the provided segments, aiming for a clear and accurate classification.
\end{itemize}

\par The specific prompt for interacting with GPT-4o is as follows:

\begin{lstlisting}
{
    "role": "system", "content":
    """
    #Objective#:
    You are an analyst. I would provide you a series of SVO pairs (the format would be mentioned later), and you're required to categorize the `"verb"` in each SVO pairs into the 8 foundational behaviour categories which we would defined later. 

    ---
    #Key Considerations#:
    - An example SVO pair may have such form:
        `"svo": [{"subject": "grandmother", "verb": "gave", "object": "a little cap made of red velvet", "receiver": "Little Red Riding Hood"}]`.
        You're required to extract the `"verb"` and categorize it based on analyzing the `"svo"`.
    - We define the 8 foundamental behaviour categories as:
        atrans: Abstract Transformation; to change an abstract relationship of an object.
        ptrans: Position Transformation; to change the location of an object.
        propel: to apply a force to an object.
        move: to move a body part.
        ingest: to take something inside an object.
        expel: to take something from inside an object and force it out.
        speak: to produce a sound.
        mental: to transfer information mentally or to create or combine thoughts.

    ---
    #Output Requirement#

    You're required to directly provide the results in JSON format based on the given example:
    ```
    [
        {
            "action": "cried",
            "category": "expel"
        },
        {
            "action": "went",
            "category": "ptrans"
        },
        {
            "action": "said",
            "category": "speak"
        },
        ...
    ]
    ```
    """
}
\end{lstlisting}

\section{Detailed information about User Study I}
\label{app:User_Study1}

\subsection{Baseline System: \textit{Cartoon Animator}}
\label{app:User_Study1_Baseline}

\begin{figure}[h]
    \centering
    \includegraphics[width=\linewidth]{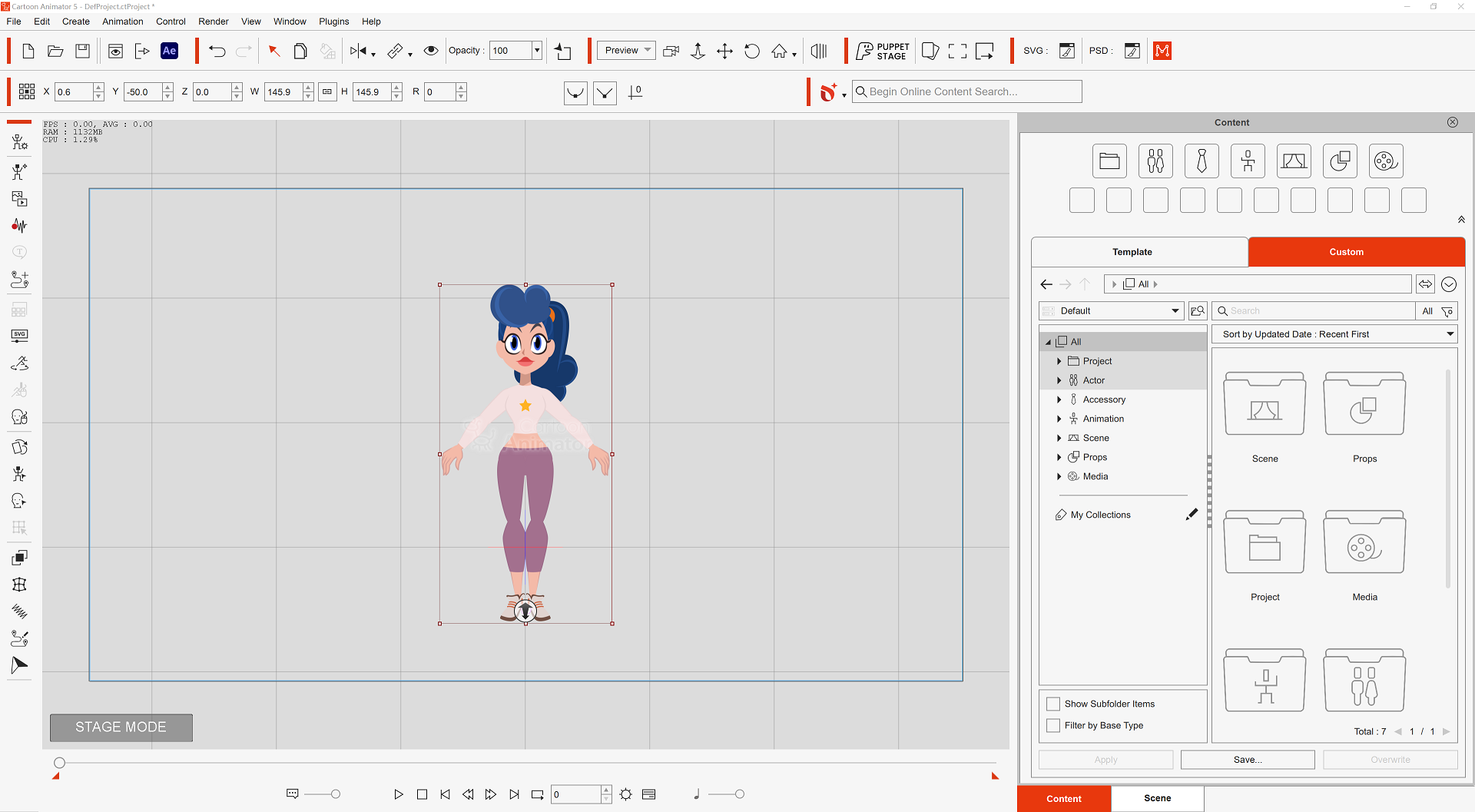}
    \caption[]{The interface of \textit{Cartoon Animator}, which introduced features such as skeleton animation editing and camera effects.}
    \label{fig:CartoonAnimator}
\end{figure}

\subsection{Population of Participants}

\begin{table}[H]
    \caption{The population of the participants of User Study I. Graphics, Animation, and Prototyping refer to the participant's experience regarding Graphics Design, Animation, and Interface Prototyping.}
    \label{tab:Study1Participants}
    \begin{tabular}{c|ccc|c|ccc}
    \hline
    \textbf{ID}  & \textbf{Graphic}  & \textbf{Animation}  & \textbf{Prototyping}  & \textbf{ID}  & \textbf{Graphic}  & \textbf{Animation}  & \textbf{Prototyping}  \\ 
    \hline
    C1  & Average    & Average    & No Proficiency & C13 & Proficient & Average    & Proficient     \\
    C2  & Average    & Beginner   & Average        & C14 & Average    & Beginner   & No Proficiency \\
    C3  & Beginner   & Beginner   & No Proficiency & C15 & Average    & Average    & Average        \\
    C4  & Beginner   & Beginner   & Beginner       & C16 & Proficient & Proficient & Average        \\
    C5  & Average    & Proficient & Proficient     & C17 & Average    & Beginner   & Beginner       \\
    C6  & Proficient & Expert     & Average        & C18 & Average    & Beginner   & No Proficiency \\
    C7  & Average    & Average    & Beginner       & C19 & Beginner   & Beginner   & No Proficiency \\
    C8  & Average    & Average    & Average        & C20 & Average    & Average    & Beginner       \\
    C9  & Average    & Beginner   & Beginner       & C21 & Proficient & Average    & Proficient     \\
    C10 & Average    & Beginner   & Average        & C22 & Average    & Beginner   & Proficient     \\
    C11 & Average    & Average    & No Proficiency & C23 & Average    & Average    & No Proficiency \\
    C12 & Proficient & Average    & Proficient     &     &            &            &               
    \end{tabular}

\end{table}

\subsection{Task 1: Create a motion comic clip based on the given story}
\par One given story segment is excerpted from ``\textit{Sleeping Beauty}''\footnote{https://www.grimmstories.com/en/grimm\_fairy-tales/sleeping\_beauty}, and its content is as follows:

\par\textit{Now the king, being desirous of saving his child even from this misfortune, gave commandment that all the spindles in his kingdom should be burnt up. The maiden grew up, adorned with all the gifts of the wise women; and she was so lovely, modest, sweet, and kind and clever, that no one who saw her could help loving her. It happened one day, she being already fifteen years old, that the king and queen rode abroad, and the maiden was left behind alone in the castle. She wandered about into all the nooks and corners, and into all the chambers and parlours, as the fancy took her, till at last she came to an old tower. She climbed the narrow winding stair which led to a little door, with a rusty key sticking out of the lock; she turned the key, and the door opened, and there in the little room sat an old woman with a spindle, diligently spinning her flax.}

\par\textit{``Good day, mother,'' said the princess, ``what are you doing?'' - ``I am spinning,'' answered the old woman, nodding her head. ``What thing is that that twists round so briskly?'' asked the maiden, and taking the spindle into her hand she began to spin; but no sooner had she touched it than the evil prophecy was fulfilled, and she pricked her finger with it. In that very moment she fell back upon the bed that stood there, and lay in a deep sleep.}

\par\textit{Many a long year afterwards there came a king's son into that country, and heard an old man tell how there should be a castle standing behind the hedge of thorns, and that there a beautiful enchanted princess named Rosamond had slept for a hundred years, and with her the king and queen, and the whole court. The old man had been told by his grandfather that many king's sons had sought to pass the thorn-hedge, but had been caught and pierced by the thorns, and had died a miserable death.}

\subsection{Task 2: Modify certain visual elements in the clip}
Here are the corresponding tasks for the aforementioned story:
\begin{itemize}
    \item \textbf{Subtask 1:} Redesign the kingdom's environment from a medieval European setting to an ancient Asian one.
    \item \textbf{Subtask 2:} Reimagine the king's clothing to a modern one.
    \item \textbf{Subtask 3:} Change the spindle into a visually distinct item of power, such as a glowing orb or a crystal, altering the scene where the princess encounters it.
\end{itemize}

\section{Detailed information about User Study II}
\label{app:User_Study2}
\subsection{Questions for \texttt{Pro-Accountant} and \texttt{DB-Accountant}}
\begin{itemize}
    \item Why the accounting materials were presented on papers? \textbf{A}. To ensure better organization and structure. \textbf{B}. To prevent any possible leakage by computer systems. \textbf{C}. To improve clarity for further revision.
    \item What unusualness did the accountant find when reviewing the documents? \textbf{A}. There was no single error. \textbf{B}. The documents were completed ahead of time. \textbf{C}. All the figures remained consistent.
    \item Why did the accountant ask for gun? \textbf{A}. He heard some unusual noise. \textbf{B}. He had to transport valuable documents securely. \textbf{C}. He found out he was being watched. 
    \item Why did Medici wanted Sunya Ekam killed? \textbf{A}. Sunya stole tons of money when making up evidences. \textbf{B}. Sunya had raw evidences of Medici's disloyalty to The Organization. \textbf{C}. The Organization did not want to keep these smart person alive.
    \item Who hired the marksmen for killing Medici? \textbf{A}. The Organization. \textbf{B}. The accountant himself. \textbf{C}. The police.
    \item What is the payment for the accountant? \textbf{A}. a large sum of money; \textbf{B}. a famous painting; \textbf{C}. an estate in Rome.
\end{itemize}

\subsection{Questions for \texttt{Pro-Spiderman} and \texttt{DB-Spiderman}}
\begin{itemize}
    \item Why don't Spiderman want Wolverine to change the costume? \textbf{A}. He thought it would make others harder to find them. \textbf{B}. He would feel awkward if only he is costumed. \textbf{C}. He believes Wolverine looks better in his original costume.
    \item How did Wolverine explain the weird costume Spiderman wore? \textbf{A}. He was costumed like this to act as a clown in a party. \textbf{B}. He was costumed like this to complete in a superhero contest before coming. \textbf{C}. He was costumed like this because he just lost a bet.
    \item Why did Wolverine feel upset when he hung over? \textbf{A}. He believed he didn't provide enough protection for Spiderman. \textbf{B}. He thought he was a good guy but broke every promise. \textbf{C}. He previously lost a battle.
    \item What led to the conflict in the bar? \textbf{A}. Wolverine was flirting with the girlfriend of the grumpy man. \textbf{B}. The man first challenged them for their costume. \textbf{C}. Wolverine accidentally spilled his drink on the man.
    \item What did Wolverine want to tell Spiderman through the analogy? \textbf{A}. Men should do their best to chase the girls. \textbf{B}. Not every fight was worthy, but people should be prepared for every fight. \textbf{C}. Some conflicts were inevitable and should be processed smoothly.
    \item Why did Wolverine Why did Wolverine call Spiderman out for a drink? \textbf{A}. Because he thought Spiderman needed advice on his future life. \textbf{B}. Because he was upset about a recent mission and wanted to vent. \textbf{C}. Because he believed Spiderman was the only one who really know him.
\end{itemize}

\end{document}